\documentclass[twocolumn]{aastex62}
\pdfoutput=1 
\usepackage{amsmath,amstext}
\usepackage[T1]{fontenc}
\usepackage{apjfonts}
\usepackage[figure,figure*]{hypcap}

\newcommand{\degree}{$^{\circ}$}

\newcommand{\hour}{$^{\rm{h}}$}
\newcommand{\minute}{$^{\rm{m}}$}
\newcommand{\second}{$^{\rm{s}}$}
\newcommand{\fsecond}{\hbox{$.\!\!^{\rm{s}}$}}

\shorttitle{}
\shortauthors{DeFelippis et al.}

\newcommand{\sofia}{SoFiA}
\newcommand{\galfahi}{GALFA-HI}
\newcommand{\panstarrs}{Pan-STARRS}

\usepackage[colorinlistoftodos]{todonotes}

\begin{document}

\title{A Correlated Search for Local Dwarf Galaxies in \galfahi\ and \panstarrs}

\author{Daniel DeFelippis}
\affiliation{Department of Astronomy, Columbia University, 550 West 120th Street, New York, NY 10027, USA}

\author{Mary Putman}
\affiliation{Department of Astronomy, Columbia University, 550 West 120th Street, New York, NY 10027, USA}

\author{Erik Tollerud}
\affiliation{Space Telescope Science Institute, 3700 San Martin Drive, Baltimore, MD 21218, USA}

\begin{abstract}
In recent years, ultrafaint dwarf (UFD) galaxies have been found through systematic searches of large optical surveys. However, the existence of Leo T, a nearby gas-rich dwarf, suggests that there could be other nearby UFDs that are optically obscured but have gas detectable at nonoptical wavelengths. With this in mind, we perform a search of the full Galactic Arecibo $L-$band Feed Array HI (\galfahi) survey, a radio survey which covers one-third of the sky at velocities $-650 < V_{\rm{LSR}} < +650 \; \rm{km} \; \rm{s^{-1}}$, for neutral hydrogen sources. We are able to probe regions of the sky at lower Galactic latitudes and smaller $|V_{LSR}|$ compared to previous explorations. We use the Source Finding Application (\sofia) on \galfahi\ and select all sources with similar properties to Leo T and other local dwarf galaxies. We find 690 dwarf galaxy candidates, one of which is particularly promising and likely a new galaxy near the Galactic plane ($b=-8^{\circ}$) that is comparable in velocity width and HI-flux to other recently discovered local volume galaxies. We find we are sensitive to Leo T-like objects out to $1 \; \rm{Mpc}$ at velocities clear from background HI emission. We check each candidate's corresponding optical fields from \panstarrs\ and fit stars drawn from isochrones, but find no evidence of stellar populations. We thus find no other Leo T-like dwarfs within $500 \; \rm{kpc}$ of the Milky Way in the one-third of the sky covered by the \galfahi\ footprint and discuss our nondetection in a cosmological context.
\end{abstract}


\section{Introduction}
\label{s:intro}

The Milky Way (MW) environment provides a unique opportunity to study galaxy formation and evolution with its observed dwarf galaxy population. The smallest dwarf galaxies are the dimmest, and also the most abundant type of galaxy in the local environment. There have been numerous searches for, and discoveries of, individual ultrafaint dwarfs (UFDs) in the past decade (e.g., \citealp{WillmanB_05,BelokurovV_06,ZuckerD_06,BelokurovV_07,IrwinM_07}). More recently, systematic searches of large optical surveys such as the Sloan Digital Sky Survey \citep[SDSS;][]{WalshS_09} and the Dark Energy Survey \citep{BechtolK_15}, as well as radio surveys such as the Galactic Arecibo L-band Feed Array (GALFA) Survey \citep{SaulD_12} and Arecibo Legacy Fast ALFA (ALFALFA) Survey \citep{AdamsE_13} have yielded even more UFD candidates.

Discoveries of local dwarfs have gone a long way in addressing conflicts between the MW's environment and cosmological simulations. For instance, the discrepancy between the number of MW satellites found observationally and in simulations first identified by \cite{KlypinA_99} -- the so-called ``missing satellites problem'' -- has been significantly reduced by simply detecting more dwarfs \citep{SimonJ_07}. Recent simulations have found satellite populations consistent with current observations \citep{Garrison-KimmelS_18}, but discrepancies between the observed MW environment and simulated environments still remain, especially in the local volume \citep{KlypinA_15}. Based on SDSS data there are potentially hundreds of faint satellites that could be revealed by deeper surveys \citep{TollerudE_08,HargisJ_14}. Though folding in data from the Dark Energy Survey lowers that estimate considerably \citep{NewtonO_18}, there may still be $\sim 100$ undiscovered satellites that are ``missing'' simply due to the incompleteness of large sky surveys, a conclusion also supported by \cite{FritzT_18}'s finding that there should be a population of undiscovered UFDs at the apocenter, thus illustrating the importance of looking for dwarfs in previously unsearched regions of the sky.

Many such unsearched regions remain that way due to various observational difficulties. Areas covered by the Galactic plane are essentially opaque in optical wavelengths, and in velocity space, Galactic emission dominates at low velocities and complicates spectral follow-up at all wavelengths. Thus far, searches for dwarf galaxies have focused on observing their stellar light at high Galactic latitude, thus limiting the area surveyed both spatially and in distance, and biasing surveys toward those galaxies with significant stellar populations.

Though first detected optically, a UFD called Leo T has been found to have recent star formation from $< 1 \; \rm{Gyr}$ ago and a sizable reservoir of neutral hydrogen gas (HI) \citep{IrwinM_07,WeiszD_14}.
Why Leo T even exists is still not clear, as most reionization models strip dwarfs of this mass of their gas and prevent them from accreting gas after $z=1$ which would be necessary for recent star formation (e.g. \citealp{RicottiM_05,RicottiM_16}). However, some models predict dwarfs could exist as gas-rich Leo T-like objects if they evolved in relative isolation from the MW \citep{RicottiM_09}. Such models thus imply there are dwarfs that can be detected by observing their gas rather than their stars, which would allow previously unobservable areas of the sky to be probed. If such dwarfs exist and have properties similar to Leo T ($M_{\rm{HI}} \sim 4.1\times10^{5} \; M_{\odot}$ and  $w_{50} \sim 17 \; \rm{km} \; \rm{s^{-1}}$ from \citealp{AdamsE_18}),
they should be detectable with high resolution and sensitivity by HI surveys within and beyond the Local Group.

With this motivation in mind, this paper attempts to find and catalog new dwarf galaxy candidates from the full \galfahi\ survey, an HI survey that covers $\approx1/3$ of the sky at unprecedented spatial and velocity resolution \citep{PeekJ_18}. We subsequently correlate the candidates with the optical \panstarrs\ survey to investigate if they have a detectable stellar population.  In Section \ref{s:methods} we describe the surveys and software used, and detail our analysis methodologies. In Section \ref{s:results} we present a catalog of dwarf galaxy candidates from our analysis, which includes a very strong candidate galaxy in the Galactic plane, and in Section \ref{s:comparison} we compare our catalog to other surveys of HI objects. Finally, we discuss the implications of our nondetections in Section \ref{s:discussion}, and summarize our results in Section \ref{s:summary}.

\section{Methods}
\label{s:methods}

\subsection{Surveys}
\label{s:surveys}

The Galactic Arecibo $L-$band Feed Array HI (\galfahi) survey is a high spatial ($4 \; \rm{arcmin}$) and spectral ($0.74 \; \rm{km} \; \rm{s^{-1}}$, smoothed) resolution survey of HI in the MW environment. It comprises 225 data cubes with 1\arcmin~pixels that cover declinations between $-1^{\circ}$ and $38^{\circ}$ across all right ascensions, and covers velocities from $-650 < V_{\rm{LSR}} < +650$ $\rm{km} \; \rm{s^{-1}}$. The survey includes a wide range of Galactic latitudes and passes through the plane of the MW twice. We utilize GALFA-HI DR2 \citep{PeekJ_18} for this study as it covers the complete 32\% of the sky with relatively uniform coverage. \cite{PeekJ_18} quote a median root mean square (rms) noise of $0.15 \; \rm{K}$ ($16 \; \rm{mJy}$ per beam), in a $1 \; \rm{km} \; \rm{s^{-1}}$ channel for DR2. We note that GALFA-HI DR1 \citep{PeekJ_11} is deeper in some areas, but only covers approximately half the sky area in a nonuniform way. See \cite{SaulD_12} for a compact cloud catalog using the DR1 data and \cite{DonovanMeyerJ_15} for an investigation of UV counterparts to these HI candidates.

To correlate objects found in \galfahi\ with optical observations, we require an optical survey with an overlapping footprint. The \panstarrs\ survey fits this need, as it covers all parts of the sky north of $-30^{\circ}$ decl. and thus fully contains the area covered by \galfahi. \panstarrs\ was run using the 1.8 m telescope at the University of Hawaii that mapped the sky in five optical and infrared bands: $g, r, i, z$, and $y$. For further details, see \cite{ChambersK_16}.

\subsection{HI Source Finding}
\label{s:sourcefinding}

We used the Source Finding Application (\sofia) developed by \cite{SerraP_15} to search through the entire \galfahi\ DR2 dataset for Leo T-like objects. \sofia's user interface contains many optional input parameters and preprocessing choices. In this section, we enumerate the steps of our analysis before, during, and after running this program, along with our rationale.
\begin{enumerate}
\item We decided to search for sources only in velocity slices where the average brightness temperature was $<1 \; \rm{K}$, to avoid the brightest Galactic emission. For most data cubes this range was typically $\sim40 \; \rm{km} \; \rm{s^{-1}}$ wide centered around $V_{\rm{LSR}} = 0 \; \rm{km} \; \rm{s^{-1}}$, but for some cubes at lower Galactic latitude the range was as high as $175 \; \rm{km} \; \rm{s^{-1}}$ (see Figure \ref{f:1Krange}). In addition, we removed the Galactic background emission from each data cube by applying an unsharp mask to each velocity slice. Without doing this, \sofia\ would often merge discrete HI blobs at low to moderate $V_{\rm{LSR}}$ with the extended Galactic emission. After experimenting with various masks and finding no major differences in the properties of the sources \sofia\ detected, we chose a mask radius of $r=30 \; \rm{arcmin}$. This radius is both larger than the expected size of a candidate galaxy, and consistent with the smoothing box size chosen by \cite{SaulD_12} for searching the \galfahi\ DR1 data.

\begin{figure}
\fig{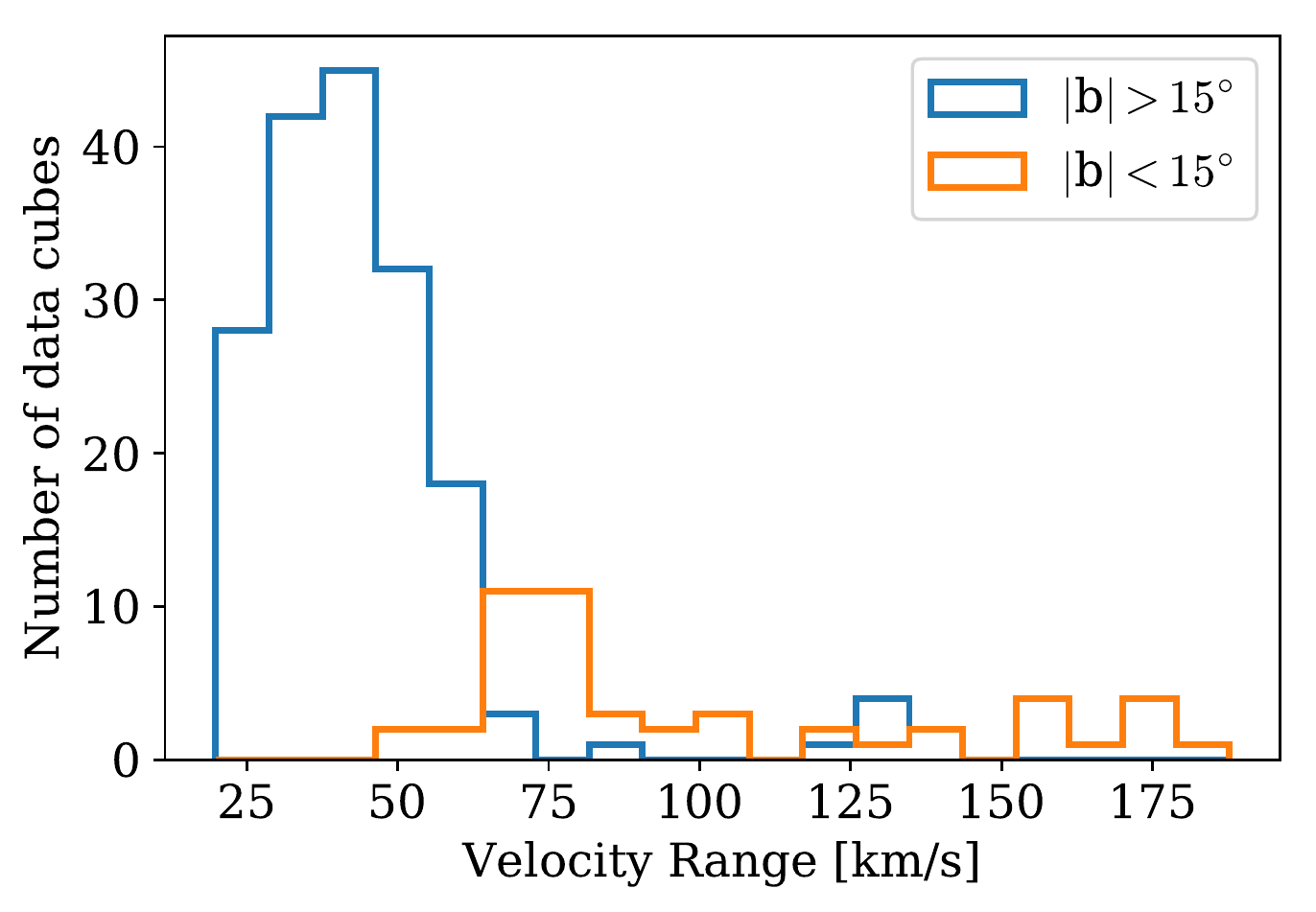}{0.5\textwidth}{}
\caption{
Distribution of Galactic emission region sizes in $\rm{km} \; \rm{s^{-1}}$ with an average brightness temperature $T_{\rm{B}} > 1 \; \rm{K}$ for all 225 data cubes, separated into regions of high and low Galactic latitude ($b$). We chose $1 \; \rm{K}$ as a compromise between being able to push to lower velocities and being able to distinguish extended Galactic emission from compact sources.
}
\vspace{0.3cm}
\label{f:1Krange}
\end{figure}

\item When running \sofia\ we turned on the noise scaling filter which normalizes the input data cube by the local noise level in each velocity slice, thus preventing faint sources from being thrown out solely because they were being compared to a less noisy background. \sofia\ measures local RMS noise by defining a box around each source, calculating the median absolute deviation of all pixels within that box that are not masked as part of the source, and multiplying by $1.4$ under the assumption of Gaussian noise (T. Westmeier 2019, personal communication). The noise level varies across the sky, largely as a function of Galactic latitude; near the plane ($|b| < 15^{\circ}$), the average RMS noise level \sofia\ calculates near detected sources is $0.21 \; \rm{K}$, while far from the plane ($b > 70^{\circ}$) it is $0.16 \; \rm{K}$.

\item For the source finding itself, we used the ``Smooth+Clip'' (S+C) finder, which smooths the data cube with user-inputted Gaussian kernels and then separates pixels that have a flux greater than some thresholds relative to the noise level. Pixels that were $\leq 1$ pixel apart in any dimension were merged into a single source. We chose a 4 pixel full width at half maximum (FWHM) for each spatial dimension to match the survey resolution, a $5$, $10$, $20$, and $30$ pixel ($3.68$, $7.36$, $14.72$, and $22.08 \; \rm{km} \; \rm{s^{-1}}$) FWHM for the velocity dimension to cover the potential ranges of velocity widths of yet undetected dwarf galaxies, and a $5\sigma$ flux threshold to reduce the probability of detecting too many spurious sources. If a source was found in any applied smoothing filter, it was then added to the catalog.

\item Once the source catalog for each cube was compiled, we removed previously known galaxies and then applied a series of cuts to remove any obviously spurious sources and tune our source list to most resemble dwarf galaxies. For reference, the properties of several recently discovered gaseous dwarf galaxies are shown in Table \ref{tab:HIgalaxies} (the definitions of the columns in the table are given in Section \ref{s:results} except for $M_*$, the stellar mass in $M_{\odot}$, and $D$, the distance to the galaxy in $\rm{Mpc}$). We set the minimum and maximum major axes of the ellipse fitted to the spatial extent of the source (called \emph{ell3s\_maj} in \sofia) to 4 and 8 pixels, respectively; the minimum and maximum velocity widths ($w_{50}$) to $10$ and $50 \; \rm{km} \; \rm{s^{-1}}$, respectively; the maximum axis ratio of the source to 1.5; and the minimum integrated signal-to-noise ratio to 20. 
\end{enumerate}

\startlongtable
\begin{deluxetable*}{lllrrrrrrrrr}
\tablecaption{
Properties of Recently Detected Local Group Galaxies with HI 
\label{tab:HIgalaxies}
}
\tablecolumns{11}
\tablenum{1}
\tablewidth{0pt}
\tablehead{
\colhead{Galaxy} & 
\colhead{R.A.} & \colhead{Decl.} & 
\colhead{$l$} & \colhead{$b$} &
\colhead{$F_{\rm{HI}}$} & 
\colhead{$M_{\rm{HI}}$} & 
\colhead{$M_{*}$} &
\colhead{$w_{50}$} & 
\colhead{$V_{\rm{LSR}}$} & 
\colhead{$D$} \\
\colhead{} & 
\colhead{(J2000)} & \colhead{(J2000)} & 
\colhead{(\degree)} & \colhead{(\degree)} &
\colhead{(${\rm Jy} \; {\rm km} \; {\rm s}^{-1}$)} & 
\colhead{($10^5 \; \rm{M_{\odot}}$)} & 
\colhead{($10^5 \; \rm{M_{\odot}}$)} &
\colhead{($\rm{km} \; \rm{s^{-1}}$)} & 
\colhead{($\rm{km} \; \rm{s^{-1}}$)} & 
\colhead{(Mpc)}
}
\startdata
Leo T & 09\hour34\minute53\fsecond4 & 17\degree03\arcmin05\arcsec & 214.85 & 43.66 & 9.9 & 4.1 & 2.0 & 17 & 34 & 0.42 \\ 
Leo P & 10\hour21\minute45\fsecond1 & 18\degree05\arcmin17\arcsec & 219.65 & 54.43 & 1.3 & 8.1 & 5.6 & 24 & 261 & 1.62 \\
Pisces A & 00\hour14\minute46\fsecond0 & 10\degree48\arcmin47\arcsec & 108.52 & -51.03 & 1.2 & 89  & 100 & 23 & 236 & 5.64 \\
Pisces B & 01\hour19\minute11\fsecond7 & 11\degree07\arcmin18\arcsec & 133.83 & -51.16 & 1.6 & 300 & 316 & 43  & 611 & 8.89 \\
\enddata
\tablecomments{
The Leo T data are from \cite{AdamsE_18}, the Leo P data are from \cite{McQuinnK_15}, and the Pisces A and B data are from \cite{TollerudE_16}.
}
\end{deluxetable*}

At the end of this process we were left with $\sim 1000$ HI sources. We did a final pruning of this source list by examining the moment maps and velocity spectra of each source by eye, and removing as candidates only those that were obviously artifacts or very close to the edge of their cube and were not already removed by the previous data cuts, or had irregular velocity spectra at high $|V_{\rm{LSR}}|$. This process resulted in a final list of 690 objects, the beginning of which is shown in Table \ref{tab:sources}. The entire list is provided in the online journal.

\subsection{Optical Correlation}
\label{s:optical}

\begin{figure}
\centering
\gridline{
\fig{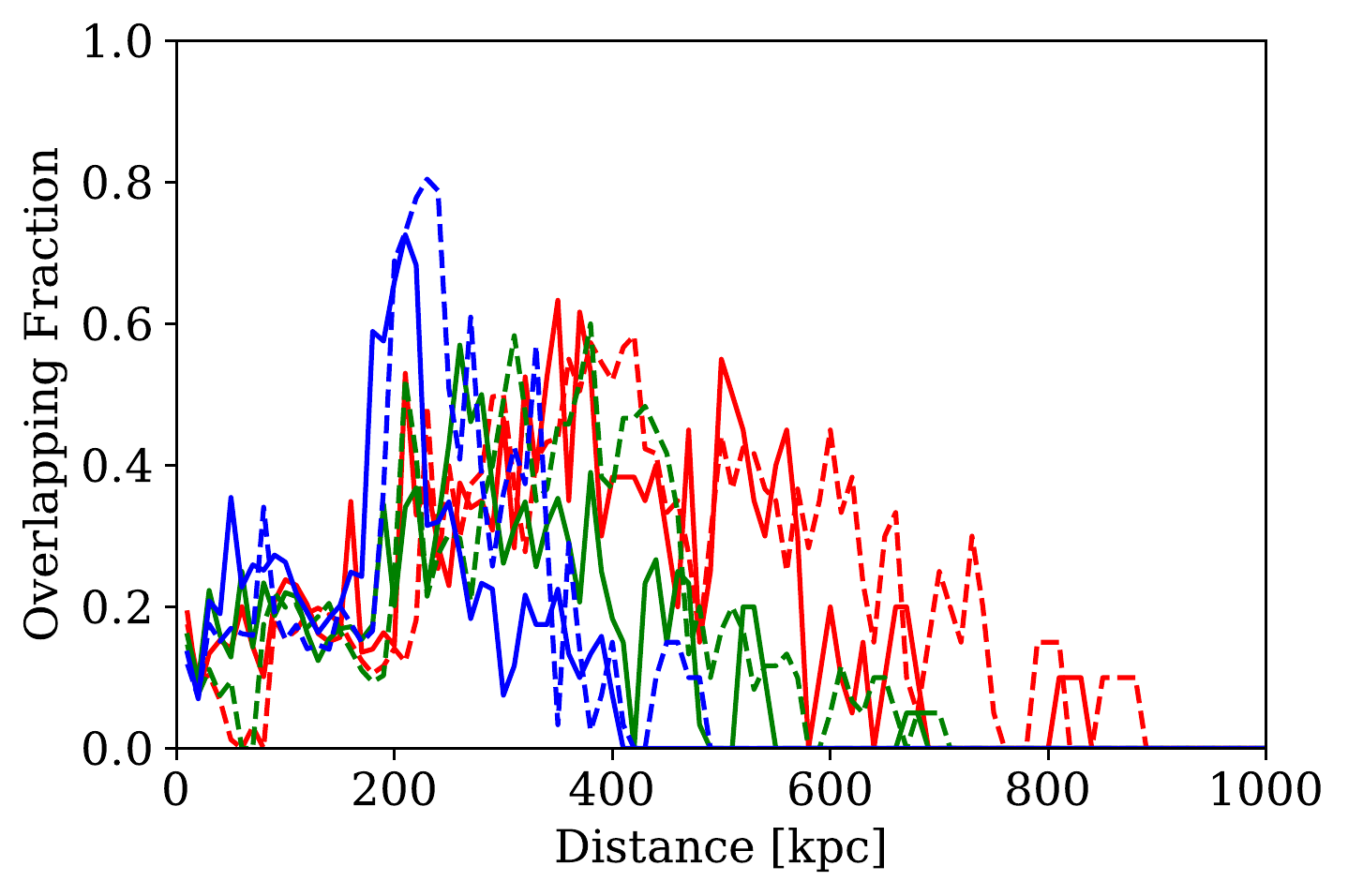}{0.25\textwidth}{}
\fig{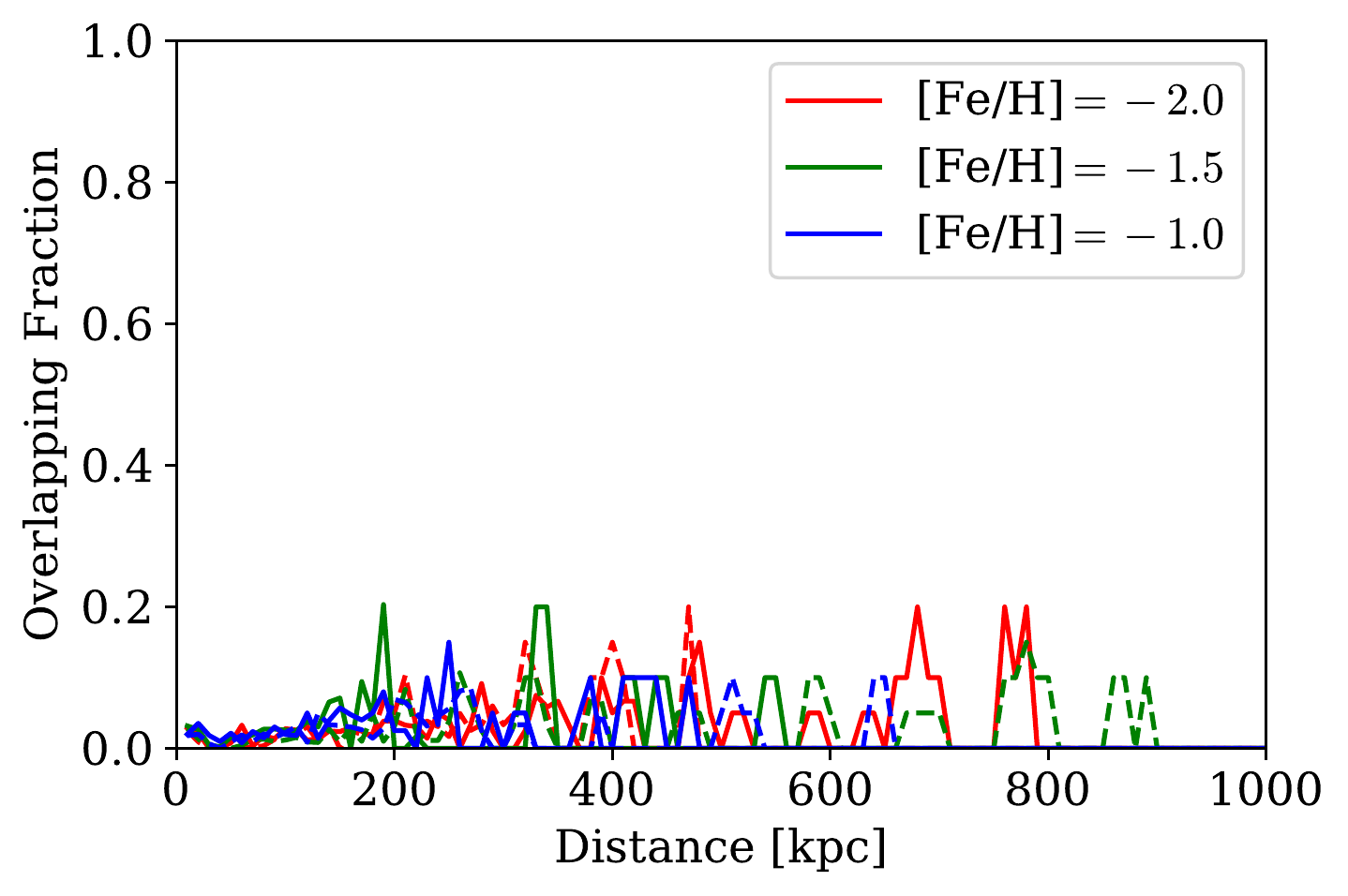}{0.25\textwidth}{}
}
\caption{
Fractions of model stellar populations that overlap with real stars in the \panstarrs\ field for Leo T, which is at a distance of $\sim 420 \; \rm{kpc}$ (left), and for the average of four control fields $10$ and $20 \; \rm{arcmin}$ away on either side of Leo T at the same Galactic latitude (right). Solid and dashed lines are $10$ and $5 \; \rm{Gyr}$ isochrones, respectively.
} 
\vspace{0.3cm}
\label{f:leotdetection}
\end{figure}

We developed and tuned an algorithm to recognize the stellar population of Leo T and other local dwarf galaxies from \panstarrs\ data, and then applied it to our HI candidates. The algorithm works as follows.
\begin{enumerate}
\item We draw a model population from a stellar isochrone of a given age and metallicity, downloaded from the CMD input form\footnote{http://stev.oapd.inaf.it/cgi-bin/cmd} \citep{BressanA_12,ChenY_14,ChenY_15,TangJ_14}. We chose from isochrones with ages of $0.5$, $2$, $5$, and $10 \; \rm{Gyr}$ and metallicities of [Fe/H] = $-1$, $-1.5$, and $-2$ to account for the large range of stellar properties of dwarfs.
\item We assign magnitudes to the model population by linearly interpolating the magnitudes and integrated IMF parameters of its isochrone, and then place it at a range of possible distances.
\item We assign magnitude and color errors based on the uncertainties provided by \panstarrs\ in the area around the source.
\item We compute the detection probability of the stars as a function of magnitude by binning the stars in the relevant \panstarrs\ field and fitting a power law. We then apply this detection probability to the model population.
\item We then compute the fraction of the model population that is within $1\sigma$ of a real star on both the magnitude and color axes. 
\item To report a detection, we require a significant peak in the overlapping fraction as a function of distance relative to a nearby control field.
\end{enumerate}
We tested this method on the resolved stellar population of the Draco Dwarf galaxy and easily recovered a distance within $10\%$ of its measured distance of $76 \; \rm{kpc}$ \citep{McConnachieA_12}. We were also able to detect Leo T with this method (Figure \ref{f:leotdetection}), though the measured distance is not very accurate. Of the three metallicities, the peak of the lowest metallicity isochrones (red lines) is closest to Leo T's known distance ($420 \; \rm{kpc}$), but the width is large. This indicates that Leo T's distance is already approaching our algorithm's detection limit when used with \panstarrs\ and that beyond that limit, further analysis would be required to confidently identify a stellar population.

\begin{figure}
\centering
\gridline{
\fig{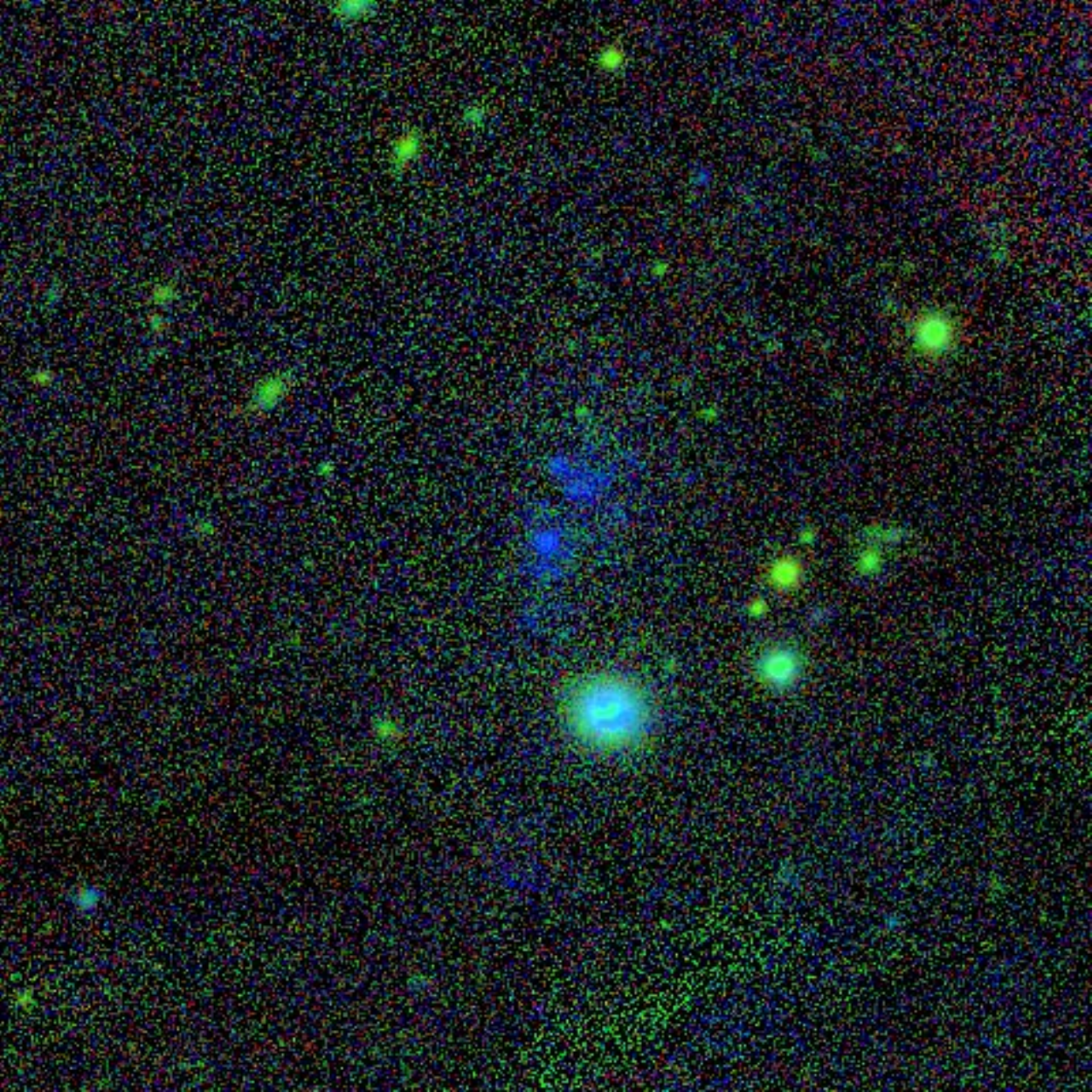}{0.25\textwidth}{}
\fig{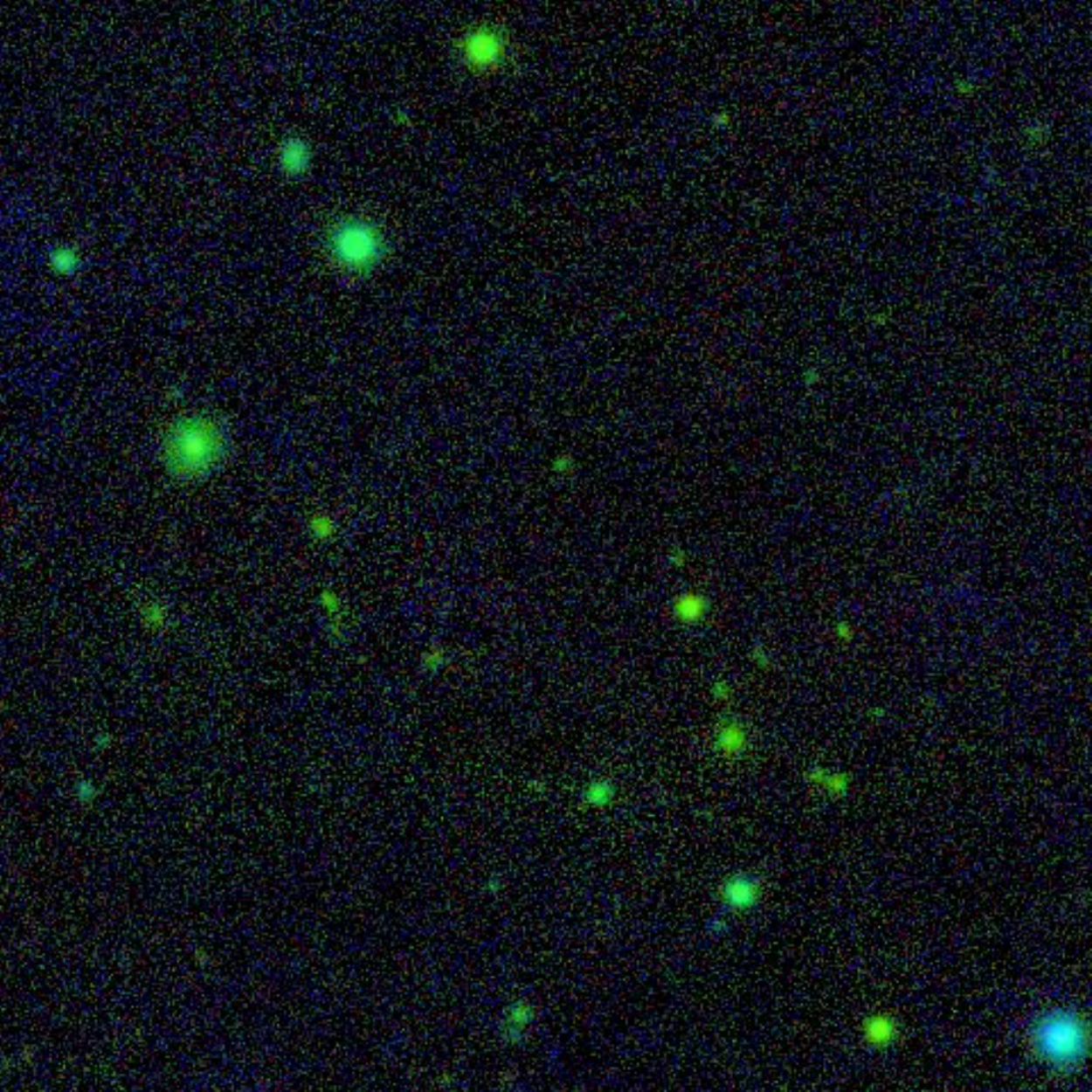}{0.25\textwidth}{}
}
\caption{
Leo P in \panstarrs\ (left) at a distance of $1.6 \; \rm{Mpc}$ which represents what we would identify as a successful detection of diffuse blue light, compared to a nearby field without a galaxy (right). Both images are $2 \; \rm{arcmin}$ across.
}
\vspace{0.3cm}
\label{f:rawimages}
\end{figure}

In addition to fitting to stellar isochrones, we also visually inspected \panstarrs\ images at each source's coordinates to identify any potential signs of a galaxy, because a galaxy's stars may be resolved or unresolved depending on its distance. We obtained uniformly scaled images by combining the $y$, $i$, and $g$ filters downloaded from the image cutout server\footnote{http://ps1images.stsci.edu/cgi-bin/ps1cutouts} with Astropy's \emph{make\_lupton\_rgb} function \citep{Astropy_18} to detect any potential stellar populations, which would show up as diffuse blue light.  Figure \ref{f:rawimages} shows a successful detection of Leo P using this method at a distance of $1.6 \; \rm{Mpc}$ \citep{McQuinnK_15}. \panstarrs\ also reveals the HI discovered galaxies Pisces A and B at $5.6$ and $8.9 \; \rm{Mpc}$ \citep{TollerudE_16}, respectively, as diffuse blue light; however, the quality of the images varies for individual sources. Leo T's stellar population does not appear as diffuse light because it is much closer and resolved.

\section{Results}
\label{s:results}

\subsection{Catalog and Sample Properties}
\label{s:catalog}

There are 690 HI galaxy candidates that were also inspected for an optical component with \panstarrs. The first 10 candidates are shown in Table \ref{tab:sources}; the entire table can be found in the online journal. The table properties are as follows.

\begin{enumerate}
\item (Source ID) A string containing (1) the Galactic longitude, (2) the Galactic latitude (both in degrees), and (3) the $V_{\rm{LSR}}$ velocity in $\rm{km} \; \rm{s^{-1}}$.
\item (R.A. and Decl.) The R.A. (hours, minutes, and seconds) and decl. (degrees, arcminutes, and arcseconds) in J2000 coordinates of the source's flux-weighted center (in other words, the ``center of flux'' of all pixels defined to be part of the source). Previous work in the ALFALFA group has found the HI positions to be accurate to within $30 \; \rm{arcsec}$ \citep{KentB_08}.
\item (Size) The major axis of the ellipse in $\rm{arcmin}$ fitted to all pixels in the source that are $\geq 3\sigma$ above the local noise level (called \emph{ell3s\_maj} in \sofia). Every such pixel is given equal weight in this calculation.
\item (S/N) The signal-to-noise ratio integrated over the entire velocity spectrum.
\item ($F_{\rm{int}}$) The flux of the source integrated over the entire velocity spectrum in ${\rm Jy} \; {\rm km} \; {\rm s}^{-1}$. This was converted from the extracted \sofia\ values in units of ${\rm K \; channel}$ using the following factor: $\frac{F_{int} \; [\rm{Jy} \; \rm{km} \; \rm{s^{-1}}]}{F_{int} \; [\rm{K} \; \rm{channel}]} = \frac{2 k_B \theta^2}{\lambda_{21}^2 }\cos(\delta) \times 0.74 \frac{\rm{km} \; \rm{s^{-1}}}{\rm{channel}} \times10^{23}$, where  $\theta = 1 \; \rm{arcmin}$ in radians, $\lambda_{21} = 21.106 \; \rm{cm}$, and $\delta$ is the source's decl.
\item ($T_{B}$) The peak brightness temperature in $\rm{K}$ (called \emph{f\_peak}, the peak flux density in \sofia).
\item ($w_{50}$) The line width at $50\%$ of the peak flux density of the source in $\rm{km} \; \rm{s^{-1}}$.
\item ($V_{\rm{LSR}}$) The local standard of rest velocity at the source's flux-weighted center in $\rm{km} \; \rm{s^{-1}}$.
\end{enumerate}

\startlongtable
\begin{deluxetable*}{llrrrrrrr}
\tablecaption{Partial Source List Sorted by Increasing Galactic Longitude
\label{tab:sources}
}
\tablecolumns{9}
\tablenum{2}
\tablewidth{0pt}
\tablehead{
\colhead{Source ID} & 
\colhead{R.A.} & \colhead{Decl.} &
\colhead{Size} & 
\colhead{S/N} & 
\colhead{$F_{\rm{int}}$} & 
\colhead{$T_B$} &
\colhead{$w_{50}$} & 
\colhead{$V_{\rm{LSR}}$} \\
\colhead{($l+b+V_{\rm{LSR}}$)} & 
\colhead{(h:m:s)} & \colhead{($^{\circ}$:$'$:$''$)} &
\colhead{(arcmin)} & 
\colhead{} & 
\colhead{(${\rm Jy} \; {\rm km} \; {\rm s}^{-1}$)} & 
\colhead{($\rm{K}$)} &
\colhead{($\rm{km} \; \rm{s^{-1}}$)} & 
\colhead{($\rm{km} \; \rm{s^{-1}}$)}
}
\startdata
000.12+75.05-034 & 13:44:18 & 18:25:34 & 5.0 & 44 & 0.83 & 0.59 & 18 & -34 \\
000.68+72.41-029 & 13:53:26 & 16:55:19 & 4.2 & 36 & 0.64 & 0.5 & 18 & -29 \\
001.51+59.47+015 & 14:35:36 & 09:02:07 & 5.0 & 40 & 1.31 & 1.03 & 20 & 15 \\
002.70+70.81-035 & 14:00:19 & 16:27:45 & 5.2 & 64 & 1.41 & 0.81 & 12 & -35 \\
003.89+68.38-062 & 14:09:29 & 15:21:07 & 7.3 & 80 & 3.14 & 0.98 & 14 & -62 \\
004.01+54.40-050 & 14:54:58 & 06:54:13 & 6.1 & 69 & 3.17 & 1.05 & 21 & -50 \\
004.27+57.23-059 & 14:46:19 & 08:46:05 & 5.8 & 51 & 1.61 & 1.0 & 13 & -59 \\
004.43+41.74-050 & 15:35:23 & -00:45:48 & 5.5 & 70 & 1.98 & 0.75 & 15 & -50 \\
004.46+58.24-068 & 14:43:19 & 09:27:47 & 4.8 & 38 & 0.93 & 0.83 & 12 & -68 \\
006.80+58.39-028 & 14:45:50 & 10:32:03 & 4.6 & 35 & 0.75 & 0.58 & 11 & -28 \\
\enddata
\tablecomments{
Uncertainty estimates for each column are given in Section \ref{s:catalog}.
}
\end{deluxetable*}

\begin{figure*}
\fig{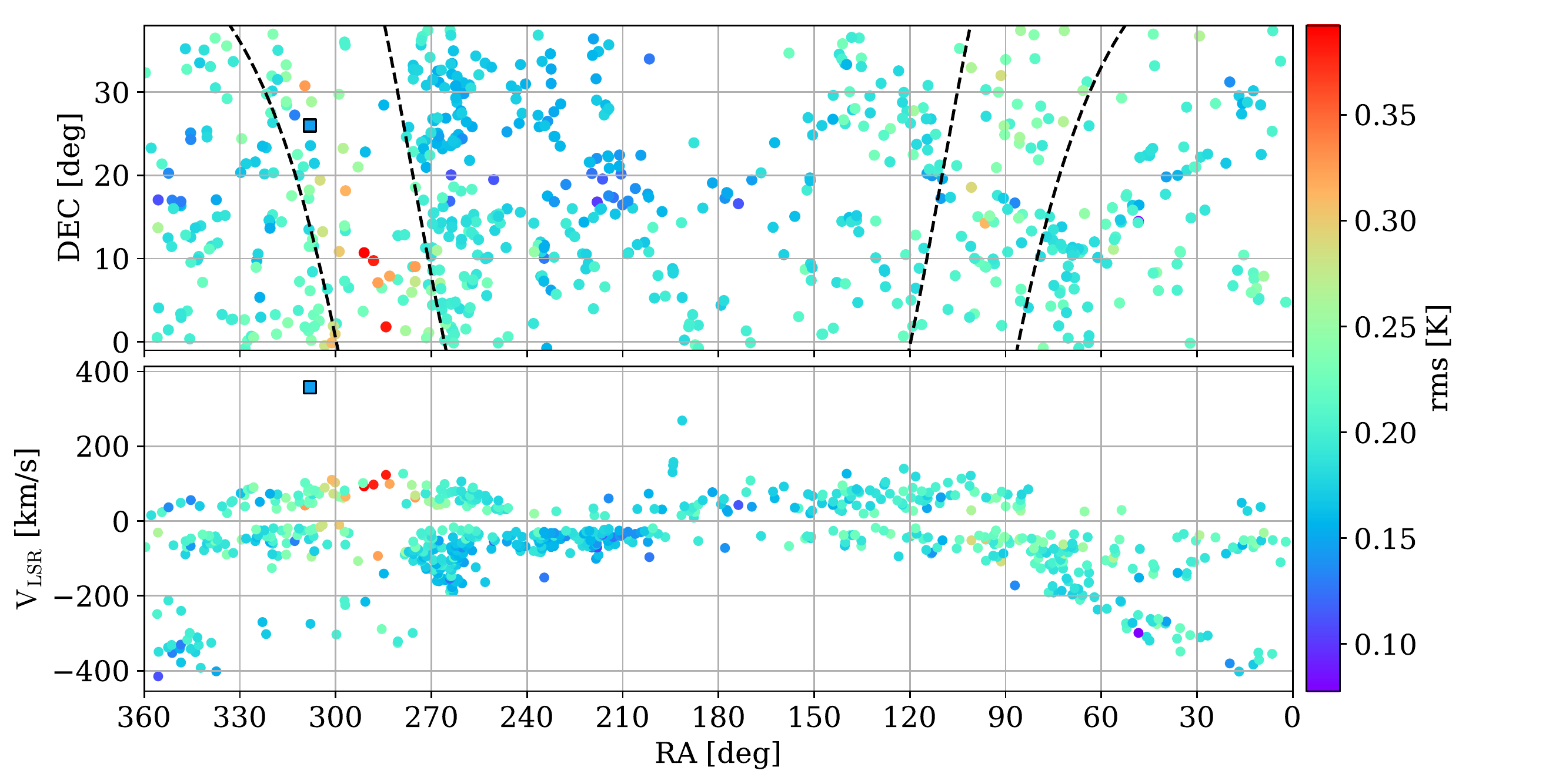}{\textwidth}{}
\caption{
Top: R.A. and Decl. positions of each source. The regions within the dashed black lines correspond to the Galactic plane in between latitudes $-15^{\circ}$ and $15^{\circ}$. Bottom: local standard of rest velocity in $\rm{km} \; \rm{s^{-1}}$ vs. R.A. for each source. Points are colored by their local RMS values calculated by \sofia. The outlined square is the local volume candidate described in Section \ref{s:candidate}.
}
\vspace{0.3cm}
\label{f:rms}
\end{figure*}

We now comment on the properties of our sample. In Figure \ref{f:rms} we see that sources are present over the entire DR2 field roughly uniformly, with a noticeable gap near R.A. $\approx 180^{\circ}$ corresponding to the North Galactic Pole. Sources are also present over a wide range of Galactic latitudes, notably including close to the Galactic plane ($|b| \lesssim 15^{\circ}$). The variation of the local noise calculated by \sofia\ when searching the data is clearly dependent on Galactic latitude (lower latitudes have systematically higher RMS values) which affects the detection limit, as discussed below. However, the distribution of the HI sources across position and velocity space shows that the noise variation is not significantly dependent on velocity. A majority ($\approx 75\%$) of sources have $|{V_{\rm{LSR}}}| < 100 \; \rm{km} \; \rm{s^{-1}}$, and we see structure at $V_{\rm{LSR}} < -200 \; \rm{km} \; \rm{s^{-1}}$ most likely associated with the Magellanic Stream. 

\begin{figure}
\fig{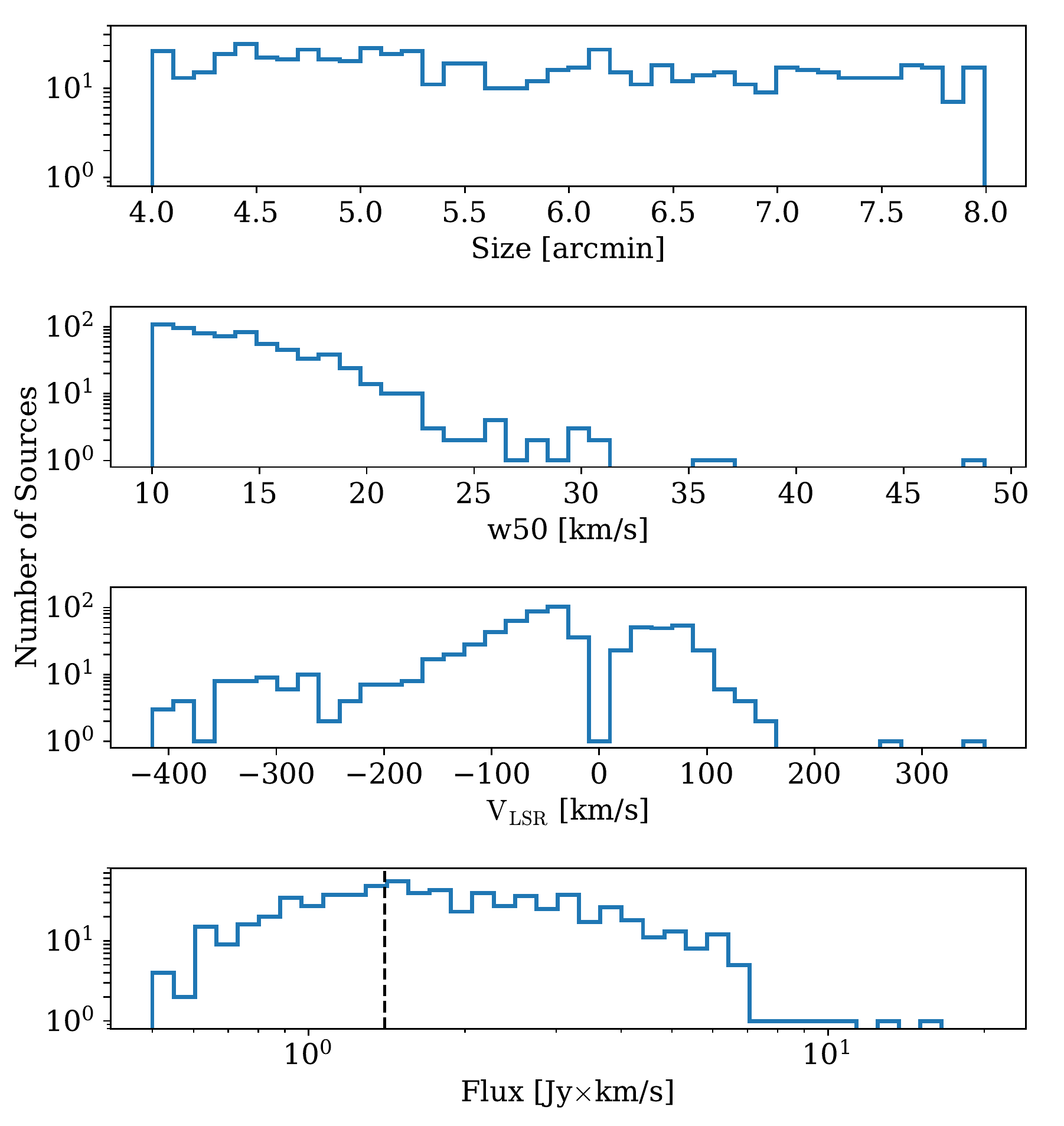}{0.5\textwidth}{}
\caption{
Distributions of angular size (top) in $\rm{arcmin}$, velocity width (upper middle) in $\rm{km} \; \rm{s^{-1}}$, local standard of rest velocity (lower middle) in $\rm{km} \; \rm{s^{-1}}$, and integrated flux (bottom) in ${\rm Jy} \; {\rm km} \; {\rm s}^{-1}$. The dashed line in the bottom panel is the estimated turnover of the flux distribution at $1.4 \; {\rm Jy} \; {\rm km} \; {\rm s}^{-1}$.
}
\label{f:properties}
\vspace{0.3cm}
\end{figure}

Figure \ref{f:properties} shows a series of properties of our detected sources. The top panel demonstrates that we are detecting objects within our desired angular size range roughly uniformly, with a slight bias to smaller objects. At a distance of $500 \; \rm{kpc}$, these sizes correspond to diameters of $0.6-1.2 \; \rm{kpc}$. The upper middle panel shows that our upper limit on the velocity width of $50 \; \rm{km} \; \rm{s^{-1}}$ was conservative; nearly all sources have a velocity width $< 20 \; \rm{km} \; \rm{s^{-1}}$. The plot also implies that the population would continue to lower velocity widths if we did not apply a cut at $w_{50} = 10 \; \rm{km} \; \rm{s^{-1}}$. 
In examining the velocity distribution (lower middle panel) we find that there is an overall bias toward objects with negative velocities; the median value of $V_{\rm{LSR}}$ is $-46 \; \rm{km} \; \rm{s^{-1}}$.

Fluxes range from $0.5$ to $15.6$ ${\rm Jy} \; {\rm km} \; {\rm s}^{-1}$, and we show the distribution of fluxes in the bottom panel of Figure \ref{f:properties}. The flux distribution turns over at $1.4 \; {\rm Jy} \; {\rm km} \; {\rm s}^{-1}$, but we note that the variability of the RMS values shown in Figure \ref{f:rms} implies that this turnover is not constant across the entire \galfahi\ field. At lower Galactic latitudes, the mean noise level is $\approx$ 10\% higher than the noise level outside of the plane. There are also known to be sources at lower fluxes than we detected in DR2 (see \cite{SaulD_12} and Figure \ref{f:allsurveys}).

We estimated errors on the properties derived by \sofia\ using injected sources, described in Section \ref{s:detection}. Position uncertainties were on the order of 15 $\rm{arcsec}$ (negligible compared to the beam size), and velocity uncertainties were on the order of 1 $\rm{km} \; \rm{s^{-1}}$, or roughly 1-2 channel spacings. Fractional uncertainties were $2-8\%$ for size measurements, $5-10\%$ for $w_{50}$ measurements, and $10-20\%$ for all other reported parameters, depending on both the distance and the velocity the injected sources were placed at. Sources at larger distances and lower velocities had higher fractional uncertainties by as much as a factor of 3.

\subsection{HI Detection Limits}
\label{s:detection}

We measure the value of the turnover in the flux distribution to be $1.4 \; {\rm Jy} \; {\rm km} \; {\rm s}^{-1}$. We convert this to a completeness limit estimate of $D_{\rm{max}} = 1.74 \times \big(\frac{M_{\rm{HI}}}{10^6 \; M_{\odot}}\big)^{1/2} \; \rm{Mpc}$ for an object with an HI-mass ($M_{\rm{HI}}$) in solar masses.
For an object with an HI-mass comparable to Leo T, this distance \sout{limit} is $\sim 1.15 \; \rm{Mpc}$. If we compute $D_{\rm{max}}$ for the galaxies in Table \ref{tab:HIgalaxies}, we find that Pisces A, B, and Leo P are all very close to their edge of detectability, and only Leo T itself is comfortably detectable. \sofia\ was indeed easily able to find Leo T in the DR2 data, but it was unable to find Pisces A, Pisces B, or Leo P. However, we were able to find Pisces A in the DR2 data cube by manually applying smoothing filters.

Due to the inherent variability of backgrounds and foregrounds across the field, we ran further tests to better quantify our estimated detection limits. We injected 3D Gaussian sources\footnote{Leo T's actual HI profiles are not exactly Gaussian \citep{AdamsE_18} so this assumption does lead to differences from Leo T's accepted mass and flux by as much as a factor of 2.} meant to replicate Leo T's accepted size and velocity width as found in \cite{AdamsE_18} (which are also consistent with the \sofia-derived size and velocity width of 6.1 $\rm{arcmin}$ and 15 $\rm{km} \; \rm{s^{-1}}$, respectively) into each cube with a peak flux corresponding to Leo T's actual distance (0.42 $\rm{Mpc}$) at three different velocities: high velocity (250 $\rm{km} \; \rm{s^{-1}}$), a low velocity defined as 15 $\rm{km} \; \rm{s^{-1}}$ greater than the boundary of the searched region defined in Section \ref{s:sourcefinding}, and Leo T's velocity (from Table \ref{tab:HIgalaxies}). For the purpose of this test, we defined a successful detection to be one where \sofia\ pulled out the injected source at its given velocity (within 15 $\rm{km} \; \rm{s^{-1}}$), position (within 5 $\rm{arcmin}$), and derived a velocity width and size within our cuts used in Section \ref{s:sourcefinding}. We recovered every injected source at high velocities, where Galactic emission is the weakest, and nearly every source at low velocities and Leo T velocities (97\% and 82\% respectively), the latter of which is lower because Leo T's velocity is often within the region of average Galactic emission $>1$K which we purposely ignored. Nevertheless, we can confidently say that \sofia\ would successfully detect Leo T at its actual distance in the vast majority of our search area.

\begin{deluxetable}{ccccc}
\tablecaption{
Detection Fraction of Sources Injected at or beyond a Distance of 500 $\rm{kpc}$
\label{tab:detectionfraction}
}
\tablecolumns{6}
\tablenum{3}
\tablewidth{0pt}
\tablehead{
\colhead{Mass (Leo T)} & 
\colhead{Distance ($\rm{Mpc}$)} & 
\colhead{High-$v$} &
\colhead{Low-$v$} &
\colhead{Leo T-$v$} 
}
\startdata
1 & 0.5 & \cellcolor{green!100}100\% & \cellcolor{green!52}76\% & \cellcolor{green!38}69\% \\
1 & 0.75 & \cellcolor{green!98}99\% & \cellcolor{red!20}40\% & \cellcolor{red!10}45\% \\
1 & 1 & \cellcolor{green!72}86\% & \cellcolor{red!78}11\% & \cellcolor{red!64}18\% \\
2 & 1 & \cellcolor{green!100}100\% & \cellcolor{red!8}46\% & \cellcolor{green!0}50\% \\
3 & 1 & \cellcolor{green!100}100\% & \cellcolor{green!36}68\% & \cellcolor{green!26}63\% \\
10 & 1 & \cellcolor{green!100}100\% & \cellcolor{green!66}83\% & \cellcolor{green!52}76\% \\
\enddata
\tablecomments{
The HI mass of Leo T is $4.1\times10^5 \; M_{\odot}$. Detection fractions are colored and shaded by how far above (green) or below (red) 50\% they are.
}
\end{deluxetable}

We next wanted to determine how accurate our detection limit of Leo T-like objects was, based on our observed flux distribution. To do this, we performed the same procedure as before with two alterations; we scaled the peak flux of the injected source to a larger distance (from 0.5 to 1 $\rm{Mpc}$) and we reduced the size of the injected source to better correspond to a more distant galaxy. At high velocities, the detection fraction dropped to its lowest value of 86\% at 1 Mpc, but the difference at low and Leo T velocities is much more drastic, as shown in the first three rows of Table \ref{tab:detectionfraction} (with Mass $= 1$ Leo T). Note, however, that a typical local group galaxy that is $1 \; \rm{Mpc}$ away has a velocity closer to our high-velocity case than either of the lower-velocity cases. In other words, the region of distance-velocity space where our detection fraction is lowest also probably contains the smallest number of dwarf galaxies. We also tried reducing the injected sources' size, which results in a $\sim 10\%$ drop in the detection fraction at lower velocities compared to the same test without changing the size, meaning the dominant reason for the overall drop is the reduction in flux caused by increasing distance. In the bottom three rows of Table \ref{tab:detectionfraction} we inject sources with successively larger multiples of Leo T's mass to determine how complete we are out to 1 $\rm{Mpc}$. We note that we would detect $\approx 50\%$ of galaxies at lower velocities with only 2 times Leo T's mass, and $\approx 66\%$ of those with 3 times Leo T's mass.\footnote{At 4 times Leo T's mass we naturally recover the same detection fractions as the first row of Table \ref{tab:detectionfraction}, because multiplying the mass of the third row by 4 is equivalent to halving the third row's distance in terms of HI flux.}

The last row of 10 times Leo T's mass is an exception to the previous rows in that the size and $w_{50}$ cuts we apply are the dominant cause for reduction in recovered sources in regions significantly contaminated by Galactic emission at low velocities. Without the size and $w_{50}$ cuts, the detection fractions are all $>92\%$. The higher level of background noise at low latitude causes \sofia\ to chop off the outer regions of the object in its size calculation.  Therefore our size cut may lead us to miss some dwarf galaxies with a range of HI masses at low latitudes and velocities.

\subsection{Optical Results}
\label{s:opticalresults}

Using the algorithm defined in Section \ref{s:optical} we also attempted to find optical counterparts for all of our sources. The vast majority of overlapping fractions were no larger than those of nearby control fields, and the few that were larger were not at all comparable to the values seen in the left panel of Figure \ref{f:leotdetection}, indicating no resolvable stellar populations. We note that Leo T is already close to the edge of detectability in \panstarrs\ and it is only $\approx400 \; \rm{kpc}$ away, so it is not necessarily surprising we were unable to see a stellar population in any of our HI sources of potentially comparable HI-masses out to $1 \; \rm{Mpc}$. We also checked the \panstarrs\ fields visually, as described in Section \ref{s:optical}, and found two potential sources with diffuse blue light (see Appendix), but otherwise nothing that indicated an unresolved stellar population similar to Leo P shown in Figure \ref{f:rawimages}.

\subsection{Local Volume Candidate}
\label{s:candidate}

\begin{figure}
\fig{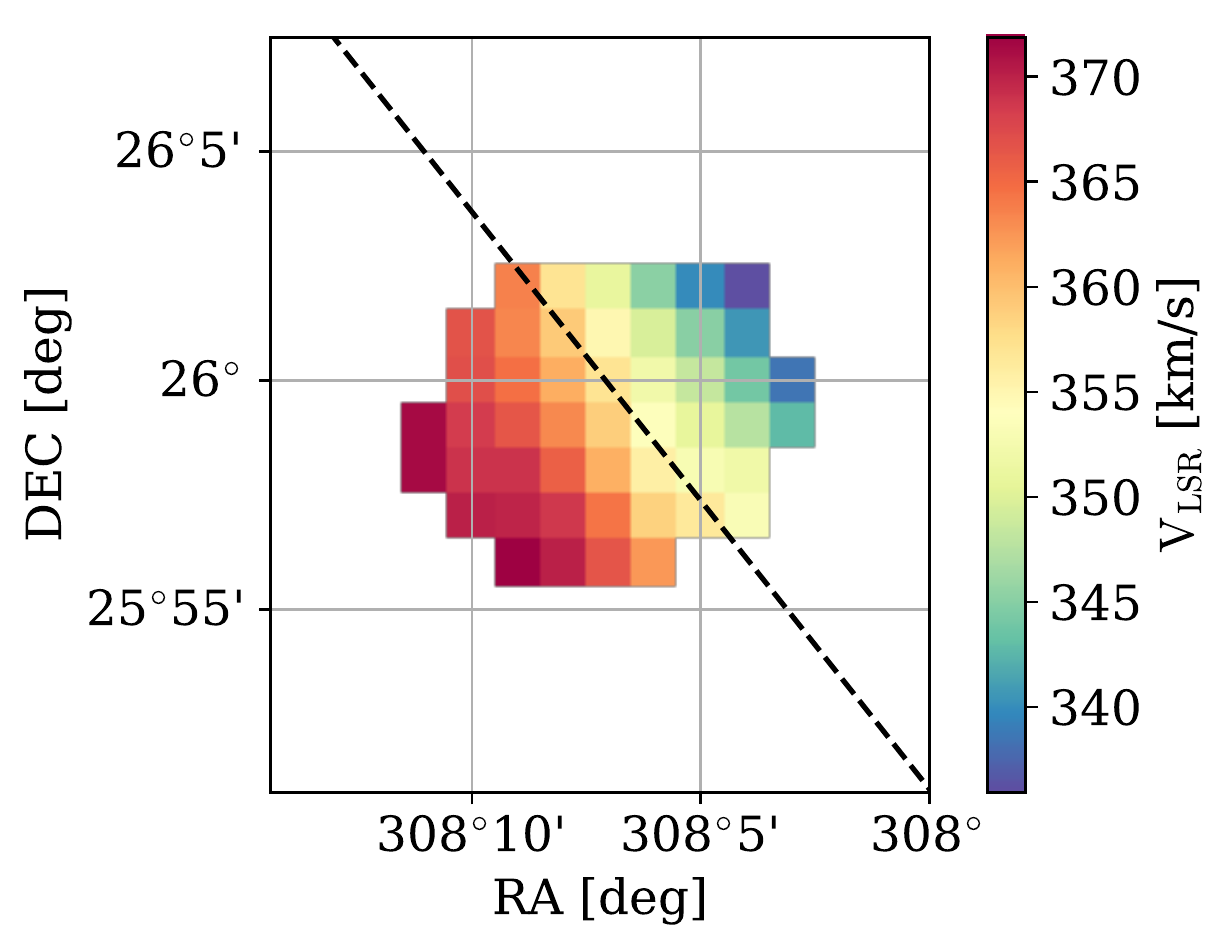}{0.5\textwidth}{}
\caption{
HI velocity map of the local volume candidate in (R.A., decl.) coordinates. Surrounding (unconnected) pixels have been masked out. The dashed line is Galactic latitude $b=-8.13^{\circ}$. 
}
\label{f:mom1}
\vspace{0.3cm}
\end{figure}

In Table \ref{tab:galaxy} we show the properties of a particularly unusual galaxy candidate that is very close to the Galactic plane ($b = -8^{\circ}$) in the constellation Vulpecula. It is an outlier in our sample in many respects. As evident in Figure \ref{f:rms}, it has the largest positive value of $V_{\rm{LSR}}$, and is the only object in our sample with $V_{\rm{LSR}} > 300 \; \rm{km} \; \rm{s^{-1}}$. It has a much larger signal-to-noise ratio than the typical value of $\approx 65$ for our source list. Its velocity width is just below our cutoff at $50 \; \rm{km} \; \rm{s^{-1}}$ and at the tail end of the distribution of velocity widths which has a mean of $\approx 15 \; \rm{km} \; \rm{s^{-1}}$ (see Figure \ref{f:properties}). We also note that it is in an area of the sky not covered by DR1 or ALFALFA. Figure \ref{f:mom1}, a velocity moment map of this source, shows evidence for a velocity gradient. Follow-up optical imaging suggests a faint optical counterpart, to be detailed in a forthcoming paper.

\begin{deluxetable}{ll}
\tablecaption{
HI properties of Local Volume candidate shown in Figure \ref{f:mom1}
\label{tab:galaxy}
}
\tablecolumns{2}
\tablenum{4}
\tablewidth{0pt}
\tablehead{
\colhead{Parameter} & 
\colhead{LV Candidate}
}
\startdata
Source ID ($l+b+V_{\rm{LSR}}$) & 067.73-08.13+358 \\
R.A. (J2000) & 20\hour32\minute26\second\\
Decl. (J2000) & 25\degree59\arcmin46\arcsec \\
Size (arcmin) & 4.4 \\
S/N & 175 \\
$F_{\rm{int}}$ (${\rm Jy} \; {\rm km} \; {\rm s}^{-1}$) & 4.97 \\
$T_B$ ($\rm{K}$) & 1.44 \\
$w_{50}$ ($\rm{km} \; \rm{s^{-1}}$) & 49 \\
$V_{\rm{LSR}}$ ($\rm{km} \; \rm{s^{-1}}$) & 358 \\
\enddata
\end{deluxetable}

\section{Comparison to Other Catalogs}
\label{s:comparison}

We find broad consistencies in the properties of our sources compared to the Compact Cloud Catalog based on \galfahi's DR1 data \citep{SaulD_12} and objects in the ALFALFA survey \citep{HaynesM_18} categorized as high velocity clouds (HVCs). For comparison to the ALFALFA catalog we used the $\alpha-100$ complete catalog available on their website\footnote{http://egg.astro.cornell.edu/alfalfa/data/}, which they state is not fully vetted but which contains Galactic sources not included in \cite{HaynesM_18}. We found sources in common between the surveys distributed uniformly across all areas of overlap on the sky. For a match, we required the difference in both R.A. and decl. to be $<5 \; \rm{arcmin}$, and the difference in the center of the velocity spectrum to be $<15 \; \rm{km} \; \rm{s^{-1}}$. We do not recover all sources from \cite{SaulD_12} or ALFALFA (68 matching objects in the former, 47 matching objects in the latter) due to differing search parameter choices and catalog methods, survey depths, and ALFALFA's lower velocity resolution. Our sky-coverage also for the first time includes the Galactic plane.

\begin{figure}
\fig{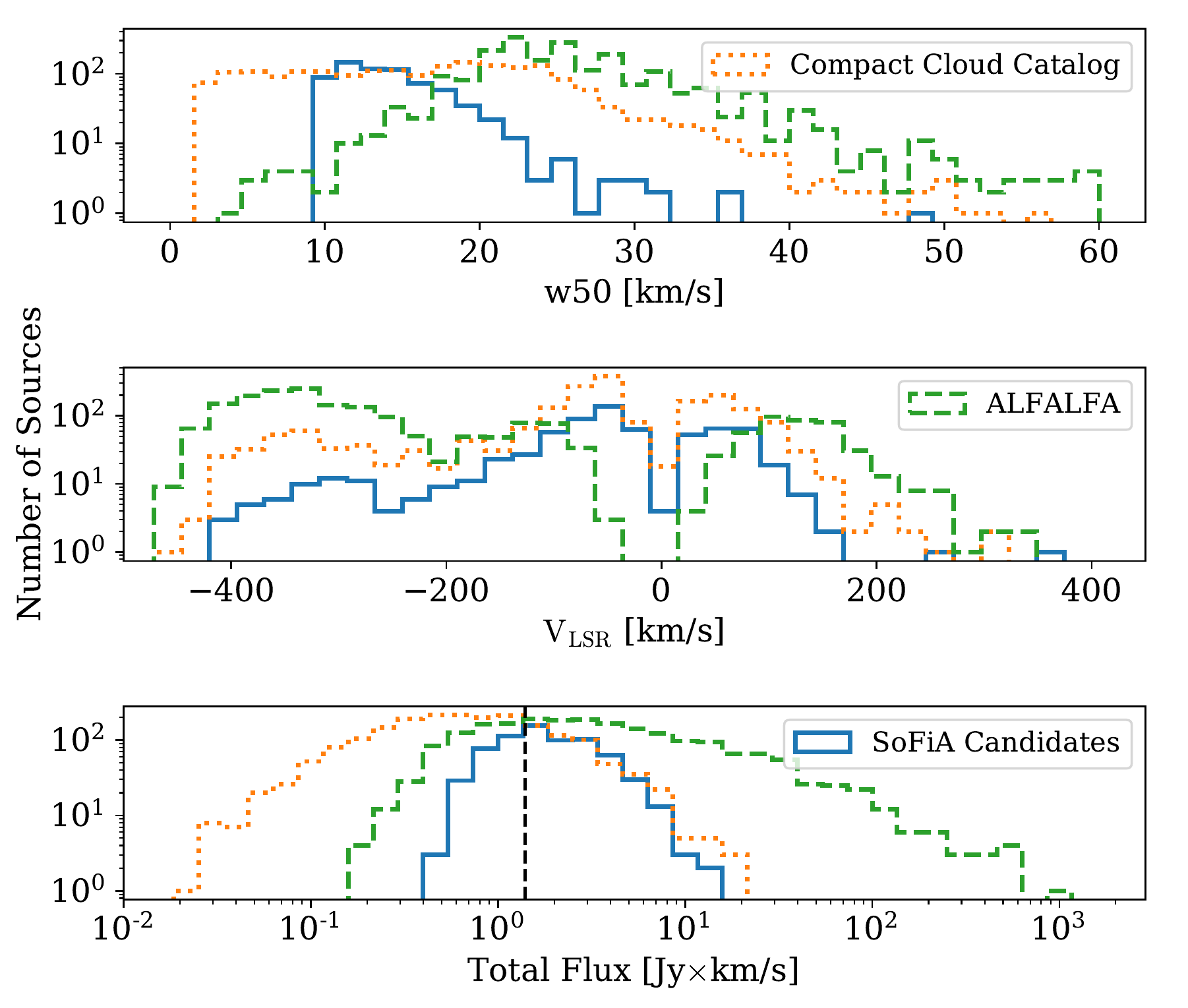}{0.5\textwidth}{}
\caption{
Distributions of velocity width (top) in  $\rm{km} \; \rm{s^{-1}}$, local standard of rest velocity (middle) in  $\rm{km} \; \rm{s^{-1}}$, and integrated flux (bottom) in ${\rm Jy} \; {\rm km} \; {\rm s}^{-1}$ for our sources (solid blue), the Compact Cloud Catalog (dotted orange), and HVCs from ALFALFA (dashed green). The dashed black line in the bottom panel is the estimated turnover of our flux distribution at $1.4 \; {\rm Jy} \; {\rm km} \; {\rm s}^{-1}$, the same as in Figure \ref{f:properties}.
}
\label{f:allsurveys}
\vspace{0.3cm}
\end{figure}

Comparisons of the same distributions in Figure \ref{f:properties} for the compact cloud and ALFALFA catalogs are shown in Figure \ref{f:allsurveys}. Our velocity width distribution (top panel) does not extend over the same range as \cite{SaulD_12}'s because we cut off sources with $w_{50} < 10 \; \rm{km} \; \rm{s^{-1}}$ and applied different masks when searching the data to focus on potential galaxy sources over clouds. Our sources extend over the velocity widths of both the cold ($\Delta V < 15 \; \rm{km} \; \rm{s^{-1}}$) and warm ($\Delta V > 15 \; \rm{km} \; \rm{s^{-1}}$) clouds defined by \cite{SaulD_12}. The difference between our catalog and the ALFALFA distribution also comes partially from the search masks we apply, but at the lower width end it is due to ALFALFA's much larger channel spacing of  $5 \; \rm{km} \; \rm{s^{-1}}$. Our sources have central velocities (middle panel) that follow the distribution of \cite{SaulD_12} very closely and have the same median $V_{\rm{LSR}} \approx -50 \; \rm{km} \; \rm{s^{-1}}$. The ALFALFA HVCs have more negative velocities (median $V_{\rm{LSR}} \approx -300 \; \rm{km} \; \rm{s^{-1}}$) most likely due to their catalog capturing many large clouds, such as those associated with the Magellanic system.

\cite{SaulD_12}'s flux distribution is essentially identical to ours down to $1.4 \; {\rm Jy} \; {\rm km} \; {\rm s}^{-1}$, though it continues to nearly two orders of magnitude fainter in flux. This is to be expected as DR1 was deeper in some areas of the sky and therefore able to pick out fainter sources. The ALFALFA catalog has a similar turnover but includes many more sources at higher fluxes, again due to the lack of size or velocity width cuts.

\begin{figure*}
\gridline{
\fig{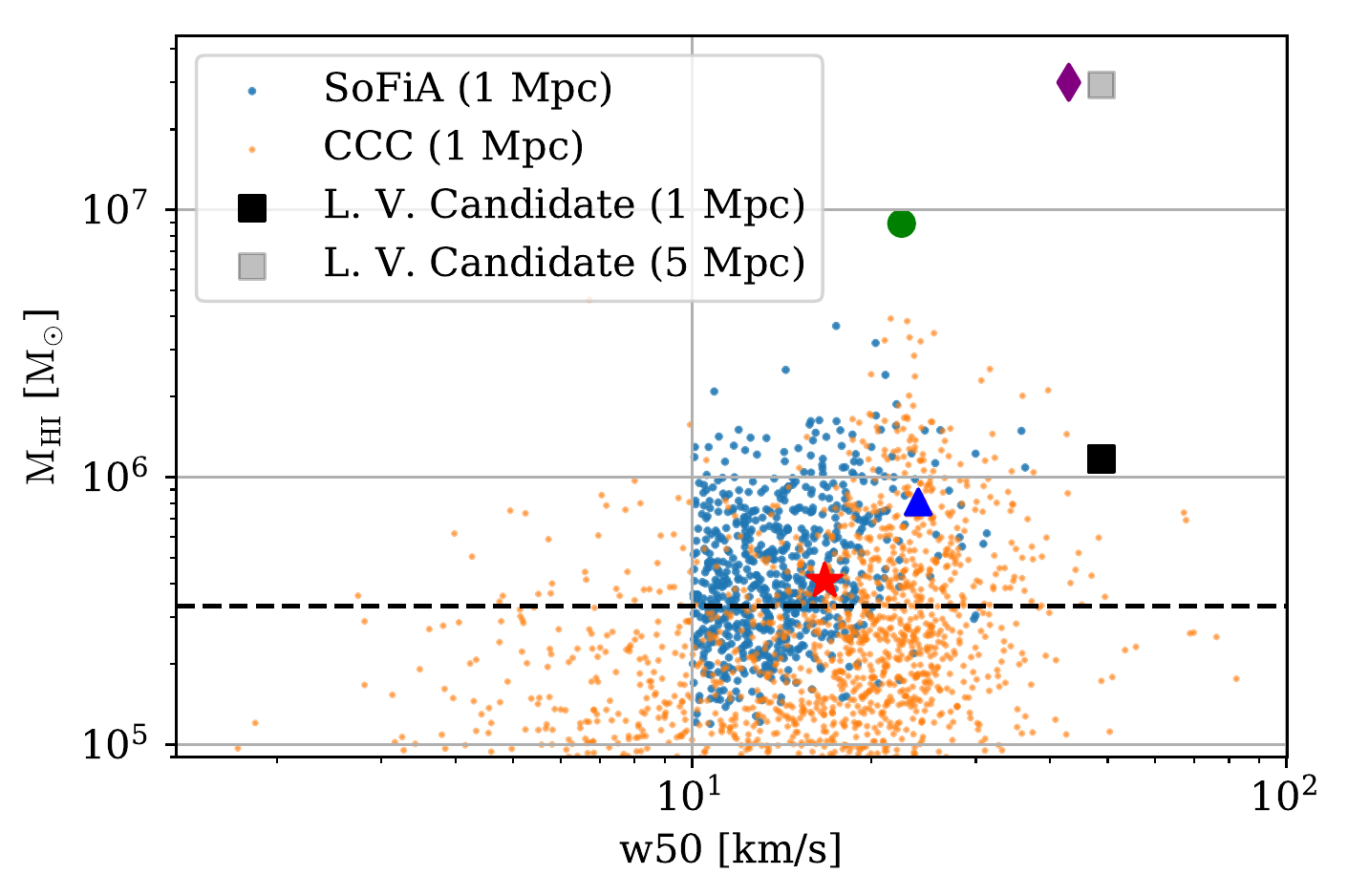}{0.5\textwidth}{}
\fig{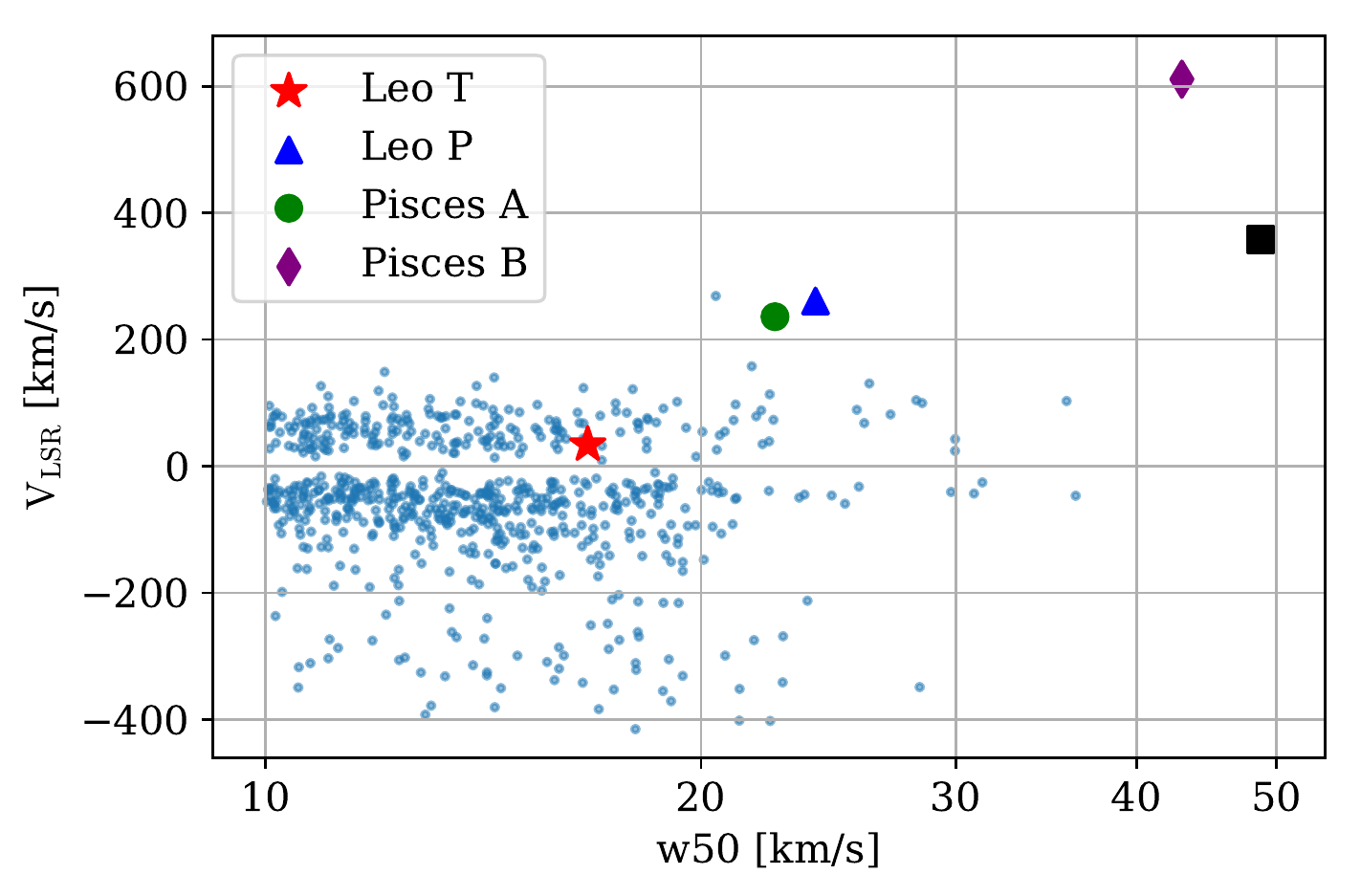}{0.5\textwidth}{}
}
\label{f:MHI_vels}
\caption{
Left: HI-mass in $M_{\odot}$ vs. velocity width ($w_{50}$) in $\rm{km} \; \rm{s^{-1}}$. The blue and orange points are our candidate sources and objects in the Compact Cloud Catalog, respectively. They are all placed at a constant distance ($1 \; \rm{Mpc}$) to derive an HI-mass. The dashed line is the HI-mass corresponding to a flux of $1.4 \; {\rm Jy} \; {\rm km} \; {\rm s}^{-1}$, also at $1 \; \rm{Mpc}$. Right: local standard of rest velocity in $\rm{km} \; \rm{s^{-1}}$ vs. $w_{50}$ for our sources. Also shown in both panels are the four galaxies in Table \ref{tab:HIgalaxies}, and our local volume candidate placed at a possible distance of $1$ and $5 \; \rm{Mpc}$. 
}
\vspace{0.3cm}
\end{figure*}

In Figure \ref{f:MHI_vels}, we directly compare properties of our candidates, the Compact Cloud Catalog, and the four reference dwarf galaxies from Table \ref{tab:HIgalaxies}. The left panel shows the HI-masses of our candidates and objects in the Compact Cloud Catalog all placed at $1 \; \rm{Mpc}$. We see that both Leo T and Leo P have comparable HI-masses compared to the derived masses of our candidates if placed at $1 \; \rm{Mpc}$. Though the right panel implies dwarf galaxies are more likely to have large positive values of $V_{\rm{LSR}}$, Leo T's existence demonstrates that local dwarfs can also have low velocities.

\section{Discussion}
\label{s:discussion}

\subsection{Local Volume Candidate}
\label{s:discussionA}

There is an observed dearth of MW satellites and local group galaxies at low Galactic latitudes (see Figure 1 of \citealp{McConnachieA_12}), presumably due to the difficulties of detecting anything in the crowded Galactic plane. If confirmed, our local volume candidate would begin to fill in the spatial distribution of nearby dwarf galaxies in regions the Galactic plane obscures. Using the galaxy candidate's measured integrated flux, we calculate it to have a HI-mass of $1.17\big(\frac{D}{1 \; \rm{Mpc}}\big)^{2}\times10^6 \; M_{\odot}$, where $D$ is its as yet undetermined distance. It has a velocity width similar to Pisces B, which was also first discovered in GALFA data, and if it also has a similar HI-mass, its HI-flux would place it at a distance of $\approx 5 \; \rm{Mpc}$ (see Figure \ref{f:MHI_vels}). This is also very close to the distance we derive assuming the candidate is in the Hubble flow ($D = V_{\rm{LSR}}/H_0$, where $H_0 = 70 \; \rm{km} \; \rm{s}^{-1} \; \rm{Mpc}^{-1}$). Using an estimate of the dynamical mass, $M_{\rm{dyn}} = 6.2 \times 10^3 a W_{50}^2 d$ (Equation (8) from \cite{AdamsE_13}), where $a$ is the angular diameter in $\rm{arcmin}$, $W_{50}$ is the velocity width in $\rm{km} \; \rm{s^{-1}}$, and $d$ is the distance $\rm{Mpc}$, we calculate a dynamical mass from our HI data of $6.5 \big(\frac{D}{1 \; \rm{Mpc}}\big) \times 10^7 \; M_{\odot}$. If $D = 5 \; \rm{Mpc}$, we derive an HI-mass-to-total mass ratio of $0.1$. We defer further analysis of this object's physical properties to a future paper with optical follow-up, which will in particular allow us to measure its distance. However, we speculate here that it is very unlikely our candidate is as close as Leo T due to its large velocity, which is inconsistent with other known galaxies in the local group \citep{McConnachieA_12}, and the absence of a Leo T-like optical image from \panstarrs: thus, it is probably not a true Leo T analog.

\subsection{Dwarf Galaxy Limits}
\label{s:discussionB}

We discuss our limits on dwarf galaxy candidates in two regimes: a galactocentric distance $<500 \; \rm{kpc}$ and a galactocentric distance $>500 \; \rm{kpc}$. We can place our strongest constraints on the existence of Leo T-like objects in the first regime, where both gas from \galfahi\ and resolved stellar populations in \panstarrs\ are detectable (see Section \ref{s:methods}). The fact that we do not see stellar populations in any of our candidates shows that there are no other Leo T-like objects closer than 500 kpc within the GALFA-HI footprint, except possibly near the Galactic plane (where the stellar population would be indistinguishable from foreground stars) or near $V_{\rm{LSR}} = 0 \; \rm{km} \; \rm{s^{-1}}$ (where we did not search to avoid bright Galactic emission). This nondetection is not completely surprising, as ram pressure stripping and other mechanisms near the MW (at $\sim 250 \; \rm{kpc}$) are expected to deplete the gas in dwarfs that reside there \citep{GrcevichJ_09,NicholsM_11,SpekkensK_14,EmerickA_16}. Nevertheless, there remains a volume of $\approx 0.17 \; \rm{Mpc^{3}}$ (the \galfahi\ sky out to $500 \; \rm{kpc}$) where we detect no additional Leo T-like objects. Our nondetection within this distance is consistent with a reionization epoch during which gas is removed from dwarf galaxies in the local group with halo masses $\lesssim 10^{8.5} \; M_{\odot}$ \citep{TollerudE_18}. In summary, an object like Leo T appears to be a rarity $<500 \; \rm{kpc}$ from the MW in the \galfahi\ footprint (which covers one-third of the sky), rather than one of many such objects. 

At distances $>500 \; \rm{kpc}$, we are unable to detect resolved stellar populations in \panstarrs, and our ability to detect Leo T-mass objects is significantly reduced at lower velocities. We can detect Leo T-like objects in HI at high velocities clear of background emission at 1 Mpc, but we only have that same level of completeness at low velocities for sources with 10 times Leo T's mass at 1 $\rm{Mpc}$. More distant objects with larger stellar populations than Leo T can be detected as diffuse blue light in \panstarrs\ (e.g. as Leo P appears in Figure \ref{f:rawimages}), and we inspected the data for these optical sources. Though we have not quantified the distance range and stellar population detectable as diffuse blue light in \panstarrs, the lack of any visual detections is consistent with all of our candidates' stellar masses being less than Leo P's stellar mass of $5.6\times10^5 \; M_{\odot}$ at a distance of $1.62 \; \rm{Mpc}$. Therefore, despite the fact that many of our candidates would be comparable in HI-mass to Leo P at distances $\gtrsim 1 \; \rm{Mpc}$ (see Figure \ref{f:MHI_vels}), the general lack of diffuse blue light is not encouraging that a large number of these are new galaxies. 
It also remains possible that some of our candidates are dark matter halos that just contain HI, as gas-rich minihalos without stars are predicted to exist around the MW \citep{RicottiM_09}.

If we consider our results in the context of the missing satellites problem \citep{KlypinA_99} as well as more recent galaxy count mismatches in the local field \citep{KlypinA_15}, we see that they support a model of strong and effective reionization that limits star formation in satellites at later times. However, the precise mechanism that operates during reionization is not yet clear. Both \cite{BrownT_14} and \cite{TollerudE_18} invoke reionization from massive stars in early galaxies to explain the current population of local dwarfs; the former by measuring ancient stellar populations, and the latter by setting a halo mass at which a dwarf cannot retain gas. Our lack of detections of Leo T-like objects strengthens this interpretation. We note though that when looking directly at star-formation histories, \cite{WeiszD_14} could not conclusively determine the effect reionization had, if any, on local dwarfs, meaning more observations are necessary to be able to distinguish between reionization models.

\begin{figure*}
\fig{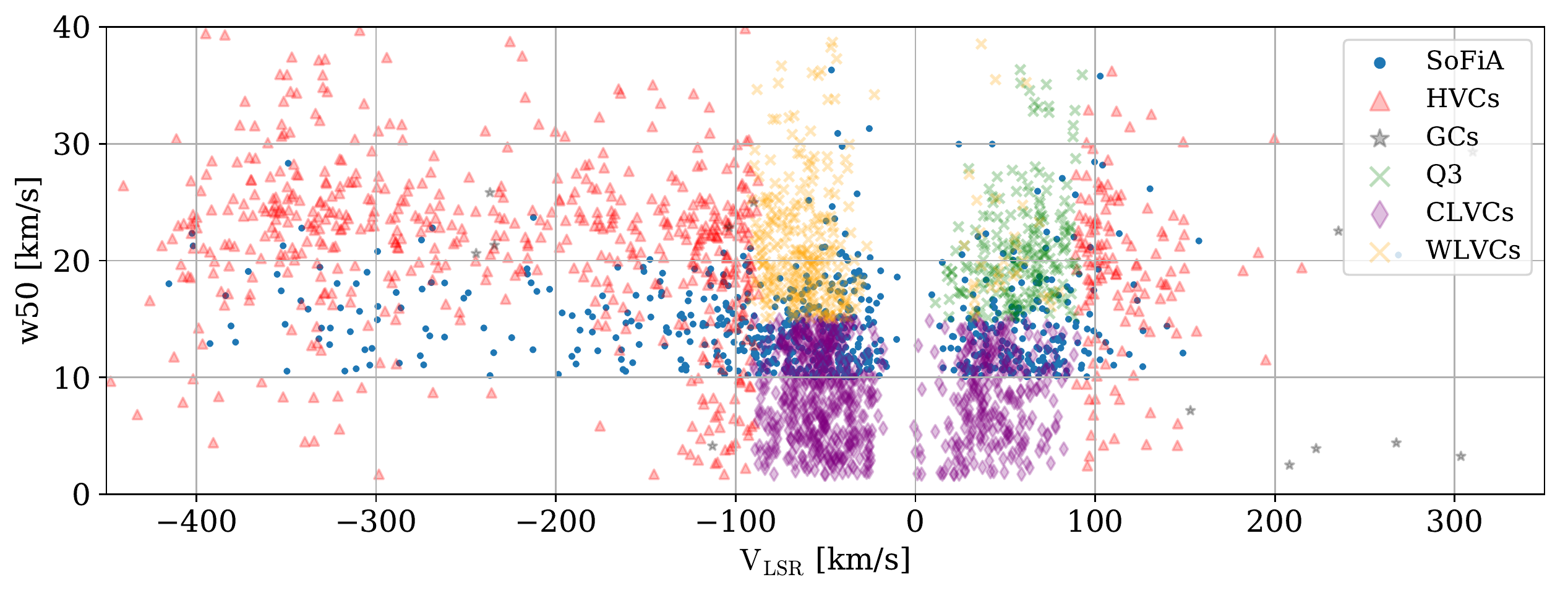}{\textwidth}{}
\label{f:CCCcompare}
\caption{
Velocity width vs. local standard of rest velocity for \sofia\ candidates and Compact Cloud Catalog objects, the latter of which is separated into HVCs which are near known complexes and have $|V_{\rm{LSR}}| > 90 \; \rm{km} \; \rm{s^{-1}}$, galaxy candidates (GCs) not near known complexes with $|V_{\rm{LSR}}| > 90 \; \rm{km} \; \rm{s^{-1}}$, cold low-velocity clouds (CLVCs) with $|V_{\rm{LSR}}| < 90 \; \rm{km} \; \rm{s^{-1}}$ and $w_{50} < 15 \; \rm{km} \; \rm{s^{-1}}$, warm low-velocity clouds (WLVCs) with $|V_{\rm{LSR}}| < 90 \; \rm{km} \; \rm{s^{-1}}$ and $w_{50} > 15 \; \rm{km} \; \rm{s^{-1}}$, and Q3 WLVCs with $0 < V_{\rm{LSR}} < 90 \; \rm{km} \; \rm{s^{-1}}$, $w_{50} > 15 \; \rm{km} \; \rm{s^{-1}}$, and $180^{\circ} < l < 270^{\circ}$.
}
\vspace{0.3cm}
\end{figure*}

Finally, we consider what other classifications for our HI sources are possible besides small galaxies. In Figure \ref{f:CCCcompare}, we compare our candidates to objects in the Compact Cloud Catalog of \cite{SaulD_12}. For the 68 matches between the two, we see a strong overlap between different types of clouds; our candidate list contains objects that overlap all types of HI sources identified in \cite{SaulD_12}, including high-velocity clouds, cold and warm low-velocity clouds, and galaxy candidates far from known HI complexes. Therefore, a plausible scenario is that most of these HI sources are a heterogeneous mixture of nearby MW structures. However, we note that it is possible that galaxies sitting at the outskirts of the local group may have small velocities (like Leo T) and could have been identified as a low-velocity cloud in \cite{SaulD_12}.

\section{Summary}
\label{s:summary}

Using Data Release 2 of \galfahi\ we performed a search for new local dwarf galaxies. We found 690 candidates, among which is an extremely promising candidate in the Galactic plane that is likely within the local volume at $V_{\rm{LSR}} = 358 \; \rm{km} \; \rm{s^{-1}}$. We quantified our completeness by injecting Leo T-like sources into each \galfahi\ data cube and measuring the fraction of sources detected by \sofia. We found we were complete out to 1 $\rm{Mpc}$ at high velocities and out to Leo T's distance at low velocities for Leo T-like dwarfs. We searched \panstarrs\ for resolved stellar populations and found none comparable to Leo T's, thus ruling out the existence of other Leo T-like dwarfs within the \galfahi\ footprint at distances $<500 \; \rm{kpc}$, except possibly at the lowest Galactic latitudes and local standard of rest velocities. We also searched for unresolved stellar populations manifesting as diffuse blue light in \panstarrs\ images, but again found no evidence of any, which limits the number of more massive dwarfs in the vicinity of the local group. We conclude that our results are consistent with strong reionization effects on the evolution of dwarf galaxies. Finally, we highlight some of our strongest candidates for potential follow-up observations in the Appendix.

\acknowledgments

We thank the entire GALFA team, and in particular Josh Peek, Yong Zheng, and Susan Clark for their help with the analysis of the \galfahi\ data and for useful discussions. We also thank Tobias Westmeier for assistance with interpreting the output from \sofia, and the anonymous referee for helpful comments. M.E.P. acknowledges support from the National Science Foundation under grant No. AST-1410800.
\software{This research made use of Astropy, a community-developed core Python package for Astronomy \citep{Astropy_18}, the open-source software tools Numpy, Matplotlib, and IPython \citep{HunterJ_07,PerezF_07,van_der_WaltS_11}, and the \sofia\ source finding pipeline \citep{SerraP_15}.}

\bibliographystyle{yahapj}
\bibliography{references.bib}

\appendix

\begin{figure}
\fig{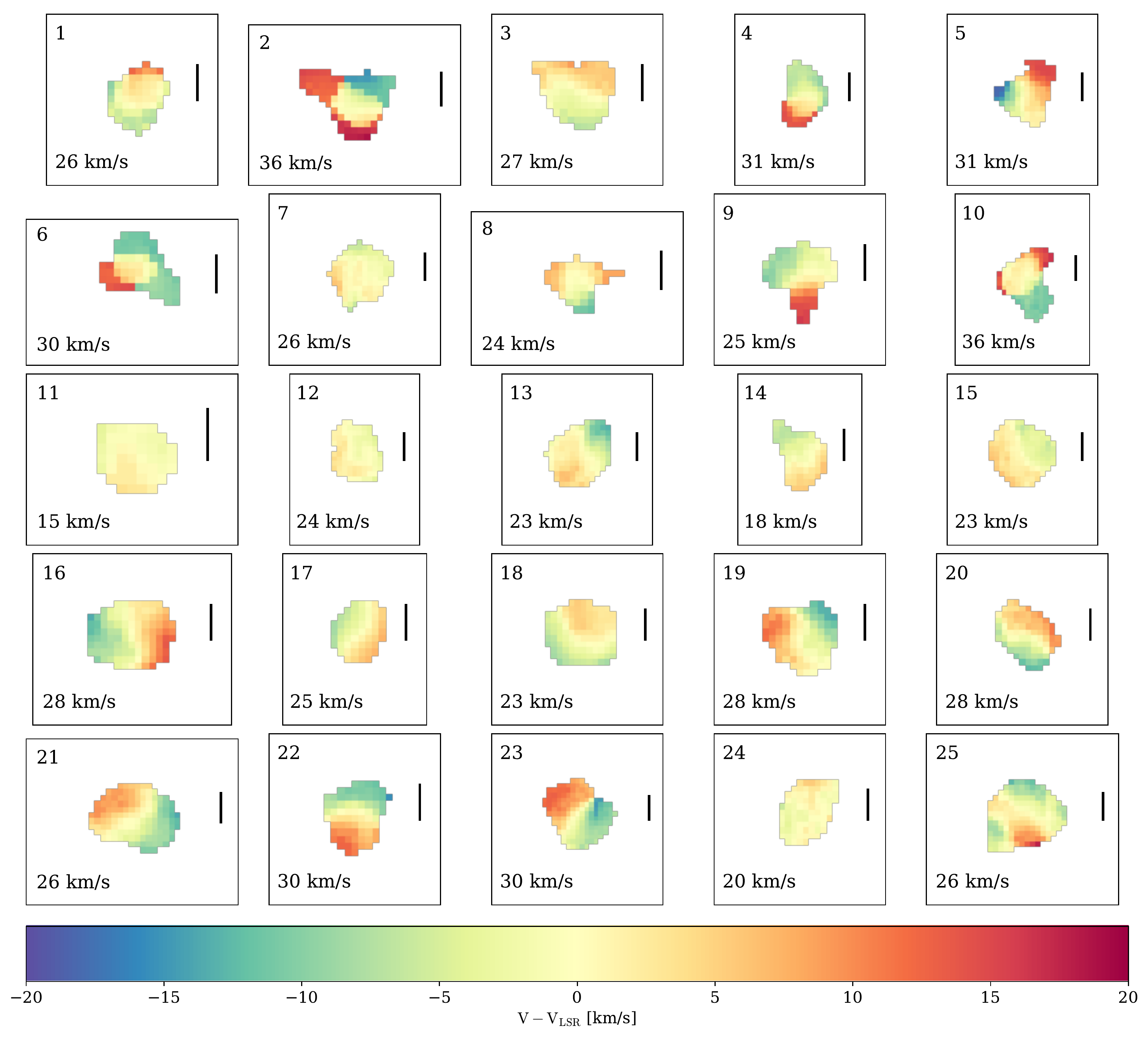}{\textwidth}{}
\label{f:appendixfig}
\caption{
HI velocity maps of the sources in Table \ref{tab:appendix_sources} in (R.A., decl.) coordinates. Red and blue colors correspond to velocities greater than and less than the systemic velocity, respectively. The numbers in the top left corners match with the entries in the first column of Table \ref{tab:appendix_sources}. The quantities in the bottom left corners are the corresponding velocity widths ($w_{50}$). The vertical black lines are each $5 \; \rm{arcmin}$ long.
}
\vspace{0.3cm}
\end{figure}

Table \ref{tab:appendix_sources} lists the most promising candidates in our catalog for potential follow-up based on their HI and/or optical properties. The sources marked with a $B$ are potentially associated with diffuse blue light offset from the center of the HI source by $\lesssim 3 \; \rm{arcmin}$. The source marked with a $V$ is isolated in velocity space as the only source other than the Local Volume candidate with $V_{\rm{LSR}} > 200 \; \rm{km} \; \rm{s^{-1}}$. All of the other sources have velocity widths larger than that of Pisces A ($w_{50} > 22.5 \; \rm{km} \; \rm{s^{-1}}$). Velocity maps of these sources are shown in Figure \ref{f:appendixfig}.

\startlongtable
\begin{deluxetable}{llrrrrrrrr}
\tablecaption{
Promising Candidates to Target for Potential Follow-up Observations, Sorted by Increasing Galactic Longitude
\label{tab:appendix_sources}
}
\tablecolumns{10}
\tablenum{5}
\tablewidth{0pt}
\tablehead{
\colhead{\#} & 
\colhead{Source ID} & 
\colhead{R.A.} & \colhead{Decl.} &
\colhead{Size} & 
\colhead{S/N} & 
\colhead{$F_{\rm{int}}$} & 
\colhead{$T_B$} &
\colhead{$w_{50}$} & 
\colhead{$V_{\rm{LSR}}$} \\
\colhead{} & 
\colhead{($l+b+V_{\rm{LSR}}$)} & 
\colhead{(h:m:s)} & \colhead{($^{\circ}$:$'$:$''$)} &
\colhead{(arcmin)} & 
\colhead{} & 
\colhead{(${\rm Jy} \; {\rm km} \; {\rm s}^{-1}$)} & 
\colhead{($\rm{K}$)} &
\colhead{($\rm{km} \; \rm{s^{-1}}$)} & 
\colhead{($\rm{km} \; \rm{s^{-1}}$)}
}
\startdata
1 & 024.86+24.05+068 & 17:11:31 & 04:06:42 & 5.2 & 57 & 2.16 & 0.91 & 26 & 68 \\
2 & 041.87-14.92+103 & 20:00:41 & 00:55:55 & 7.2 & 104 & 6.32 & 1.68 & 36 & 103 \\
3 & 046.10-09.25+082 & 19:48:46 & 07:17:17 & 6.1 & 82 & 3.78 & 0.99 & 27 & 82 \\
4 & 047.56+27.33-043 & 17:32:58 & 24:08:54 & 6.5 & 80 & 2.39 & 0.89 & 31 & -43 \\
5 & 050.68-19.13-026 & 20:32:08 & 06:08:02 & 6.6 & 60 & 2.62 & 1.05 & 31 & -26 \\
6 & 052.97+33.93-041 & 17:10:32 & 30:29:55 & 5.8 & 58 & 1.25 & 1.02 & 30 & -41 \\
7 & 063.19+22.95-032 & 18:12:46 & 36:12:26 & 6.0 & 71 & 2.59 & 0.91 & 26 & -32 \\
8 & 063.94+24.91-045 & 18:04:25 & 37:26:22 & 4.1 & 43 & 0.93 & 0.88 & 24 & -45 \\
9 & 073.70-11.75-059 & 21:01:30 & 28:25:59 & 5.4 & 56 & 1.8 & 0.93 & 25 & -59 \\
10 & 083.17-44.66-047 & 23:01:32 & 09:33:36 & 7.4 & 87 & 4.61 & 1.06 & 36 & -47 \\
11 & $^B$084.85-40.49-351 & 22:55:57 & 13:41:39 & 4.1 & 45 & 0.97 & 0.88 & 15 & -351 \\
12 & 085.14-55.33-212 & 23:29:48 & 01:26:26 & 6.2 & 94 & 4.17 & 1.05 & 24 & -212 \\
13 & 092.06-40.55-341 & 23:15:28 & 16:30:01 & 7.2 & 96 & 3.62 & 0.85 & 23 & -341 \\
14 & $^B$110.15-29.23-069 & 23:58:36 & 32:19:32 & 6.1 & 46 & 1.27 & 0.88 & 18 & -69 \\
15 & 154.71-36.35-269 & 02:38:37 & 19:49:10 & 6.6 & 81 & 2.82 & 0.71 & 23 & -269 \\
16 & 155.71-46.33-349 & 02:20:54 & 10:45:56 & 6.7 & 70 & 3.33 & 1.18 & 28 & -349 \\
17 & 197.33+00.91-046 & 06:25:43 & 14:16:26 & 4.8 & 119 & 6.34 & 1.9 & 25 & -46 \\
18 & 201.91-01.75-050 & 06:24:51 & 08:59:15 & 6.5 & 109 & 5.2 & 1.13 & 23 & -50 \\
19 & 216.76+26.13+100 & 08:31:45 & 08:28:54 & 5.7 & 64 & 2.33 & 0.84 & 28 & 100 \\
20 & 218.47+25.07+104 & 08:30:43 & 06:37:48 & 6.1 & 61 & 2.53 & 0.79 & 28 & 104 \\
21 & 218.81+15.33+089 & 07:56:36 & 01:52:08 & 7.7 & 91 & 4.79 & 1.1 & 26 & 89 \\
22 & 239.80+69.44+043 & 11:35:06 & 16:35:54 & 5.5 & 63 & 1.29 & 0.47 & 30 & 43 \\
23 & 291.07+62.01+024 & 12:29:17 & -00:20:50 & 7.4 & 87 & 5.2 & 1.28 & 30 & 24 \\
24 & $^V$299.01+68.15+269 & 12:45:35 & 05:19:27 & 6.1 & 50 & 1.79 & 0.7 & 20 & 269 \\
25 & 307.78+71.61+131 & 12:57:37 & 08:47:58 & 7.6 & 112 & 6.36 & 1.41 & 26 & 131 \\
\enddata

\end{deluxetable}

\end{document}